\documentclass[pra,reprint,superscriptaddress,amsmath,amssymb,aps,floatfix]{revtex4-1}
\usepackage[utf8]{inputenc}
\usepackage{hyperref}
\usepackage{graphicx}
\usepackage{comment}
\usepackage{braket}

\usepackage{color}
\usepackage[normalem]{ulem} 	
\definecolor{myblue}{rgb}{0.4, 0.3, 0.7}
\definecolor{purple}{rgb}{0.63,0,1}

\definecolor{dark-green}{rgb}{0,0.4,0.1}
\definecolor{dark-gray}{rgb}{0.4,0.4,0.4}

\definecolor{pink}{rgb}{1,0,0.9}

\newcommand{\subfigref}[2]{\hyperref[#1]{\ref*{#1}(#2)}}
\newcommand{\subfigrefs}[3]{\subfigref{#1}{#2}--\subfigref{#1}{#3}}

\begin{document}
\title{Continuous matrix-product states in inhomogeneous systems with long-range interactions}
\author{I.V. Lukin}
\email{illya.lukin11@gmail.com}
\affiliation{Karazin Kharkiv National University, Svobody Square 4, 61022 Kharkiv, Ukraine}

\author{A.G. Sotnikov}
\email{a\_sotnikov@kipt.kharkov.ua}
\affiliation{Karazin Kharkiv National University, Svobody Square 4, 61022 Kharkiv, Ukraine}
\affiliation{Akhiezer Institute for Theoretical Physics, NSC KIPT, Akademichna 1, 61108 Kharkiv, Ukraine}

\date{\today}

\begin{abstract}
We develop the continuous matrix-product states approach for description of inhomogeneous one-dimensional quantum systems with long-range interactions.
The method is applied to the exactly-solvable Calogero-Moser model. We show the high accuracy of reproducing the ground-state properties of the many-body system and discuss potential errors that can originate from the approximation of the nonlocal interaction potentials with singularities. 
\end{abstract}

\maketitle

\section{Introduction}
Over the last decades, the density-matrix renormalization group (DMRG) approach has become the method of choice in studies of gapped local one-dimensional systems on the lattice~\cite{Schollwoeck2011TheDR}. DMRG is a variational method, which represents the ground-state wave function on the one-dimensional lattice as a peculiar tensor-network structure --- a matrix product state (MPS). The success of this variational ansatz is based on the effective encoding of the entanglement structure of the ground state~\cite{Eisert2010AreaLF}. 

In the following years, the MPS approach was generalized, in particular, to critical systems with the multi-scale entanglement renormalization ansatz (MERA)~\cite{MERA}, two-dimensional systems with the projected entangled pair-state approaches~\cite{Verstraete2004RenormalizationAF}, and to description of the real-time dynamics. 
Another direction of the MPS development were continuous systems.
There, one can either study a continuous system on the lattice and extract results in the continuous limit via a certain form of scaling analysis~\cite{MultigridMethods, Dutta2021DensityMatrixRG, cMPSfinegrain, Banuls_2013} or employ the continuous generalization of the matrix product states approach (cMPS)~\cite{cMPSIntro, Haegeman2013CalculusOC}. 
Note that one can also apply the hybrid methods~\cite{PhysRevB.98.195105}, which rely on both the lattice fine graining and cMPS. 
Until recently, most of the cMPS studies were focused on various aspects of translationally-invariant systems both with short-range~\cite{cMPSfinegrain,TransInvcMPS1, Vanderstraeten2019TangentspaceMF,SciPostPhys.3.1.006} and long-range interactions~\cite{PhysRevB.92.115107} (including periodic boundary conditions) or on generalizations to relativistic systems~\cite{PhysRevD.104.096007}. 
Continuous tensor networks were also generalized for the studies of time dynamics~\cite{TimeDynamics}, high-dimensional systems~\cite{Jennings_2015, cPEPS, cPEPS2}, continuous MERA~\cite{cMERA}; 
these were also successfully applied to the finite-temperature simulation of lattice systems~\cite{cMPO}, the relation to continuous measurements~\cite{Jennings_2015, PhysRevLett.105.260401}, as well as to open quantum systems~\cite{OpenSystems, Garrahan_2016}.  

Recently, several cMPS-related methods were also suggested to describe quantum many-body systems with no translational invariance. 
They rely on different approximations for the matrices that parametrize cMPS by using splines~\cite{Ganahl2017ContinuousMP} or finite elements~\cite{PhysRevLett.128.020501}
with the succeeding employment of the steepest gradient descent methods for the parametrized wave functions. 
In this paper, we generalize the method~\cite{PhysRevLett.128.020501} to inhomogeneous systems with long-range interactions (including singular potentials)
and benchmark it on the corresponding exactly solvable model.

\section{Method}\label{sec:method}
For definiteness and simplicity, let us focus on systems consisting of interacting bosons (the generalization to fermions and multicomponent gases can be performed along the lines of Refs.~\cite{cMPSFermions, Mixtures, cMPSmixtures, BBMixtures, BFMixtures}) on a finite space interval $x \in [0,L]$. 
Bosonic particles are characterized by the creation and annihilation operators with the conventional commutation relations and related to the field operators $\psi^{\dagger}(x)$ and $\psi(x)$, respectively. The cMPS variational ansatz can be expressed as follows:
\begin{equation}
    |Q,R\rangle = \langle\nu_{L}|  P \exp{
    \int_{0}^{L}dx [Q(x)+R(x)\psi^{\dagger}(x)]}   |\nu_{R}\rangle
    \ket{0},
\end{equation}
where $R(x)$ and $Q(x)$ are the coordinate-dependent matrices of dimension $D$, 
$\nu_{L}$ and $\nu_{R}$ are the $D$-dimensional vectors,  $P\exp(\ldots)$ is the path-ordered exponent, 
and $\ket{0}$ is the vacuum state. 
$Q$, $R$, $\nu_{L}$, and $\nu_{R}$ are the variational parameters we aim to optimize. 
To this end, we employ the parametrization of general matrices $Q(x)$ and $R(x)$ from Ref.~\cite{PhysRevLett.128.020501}.
We introduce a mesh grid $[0, x_{1}, ...,x_{i},...,L]$ on the interval $[0, L]$ and define values of $R$ and $Q$ on the nodes of the grid as $R(x_{i}) = R_{i}$ and $Q(x_{i})=Q_{i}$. 
In the spatial interval between the two nearest-neighbor nodes $x_{j}$ and $x_{j+1}$, we use the linear interpolation
\begin{equation}
    R(x) = R_{i} + (R_{i+1}-R_{i}) \left( \frac{x - x_{i}}{x_{i+1} - x_{i}} \right).
\end{equation}
In this sense, $\nu_{L}$, $\nu_{R}$, $Q_{i}$, and $R_{i}$ constitute now a finite set of variational parameters.

Next, let us turn to one-dimensional quantum many-body systems with long-range two-body interactions. 
The corresponding Hamiltonian can be written in the following form:
\begin{multline}\label{Hamiltonian}
        H = \int_{0}^{L}\left\{ \frac{1}{2}\frac{d\psi^{\dagger}(x)}{dx} \frac{d\psi(x)}{dx} + [V(x)-\mu]\psi^{\dagger}(x) \psi(x) \right.
        \\
        + g \psi^{\dagger}(x) \psi^{\dagger}(x) \psi(x) \psi(x) \Bigr\} dx
        \\
        + \int_{0}^{L} dx \int_{x}^{L} dy U(y,x) \psi^{\dagger}(x) \psi^{\dagger}(y) \psi(x) \psi(y),
\end{multline}
where $V(x)$ is the external potential, $U(y,x)$ is the two-body interaction potential, $g$ is coupling constant of the local two-body interaction, and $\mu$ is the chemical potential, which controls the number of particles in the system under study.
At the moment, we do not specify the form of the nonlocal two-body interaction potential $U(x,y)$. 

To obtain the variational cMPS wave function for the Hamiltonian~\eqref{Hamiltonian}, it is necessary to compute the expectation value of the energy operator and the corresponding gradients.
Following Ref.~\cite{PhysRevLett.128.020501}, we introduce the matrices $\sigma_{L}(x)$ and $\sigma_{R}(x)$ of the size $D\times D$, which describe the wave-function density matrices to the left and to the right sides from the point~$x$, respectively. 
In terms of the wave functions corresponding to these density matrices, the expectation values of operators can be computed as
\begin{equation}
    \langle O \rangle(x) = \frac{\langle \sigma_{L}(x)|O(R,Q)|\sigma_{R}(x)\rangle}{\langle \sigma_{L}(x)|\sigma_{R}(x)\rangle},
\end{equation}
where $O(R,Q)$ is a matrix of the size $D^{2}\times D^{2}$ constructed in terms of the matrices $R(x)$ and $Q(x)$. For the physical operators such as the kinetic energy or particle density, we have the following mapping rules for the matrices $O(R,Q)$ (for derivation, see, e.g., Ref.~\cite{Haegeman2013CalculusOC}):
\begin{eqnarray}\label{Rules}
    \psi^{\dagger}(x)\psi(x) 
    &\to& R(x) \otimes \overline{R(x)} ,
    \\
    \psi^{\dagger}(x)\psi^{\dagger}(x)\psi(x)\psi(x) 
    &\to& R(x)^{2} \otimes \overline{R(x)}^{2} ,
    \\
    \frac{d\psi^{\dagger}(x)}{dx} \frac{d\psi(x)}{dx}
    &\to& DR(x) \otimes \overline{DR(x)},
\end{eqnarray}
where
$A\otimes\overline{A}$ means the Kronecker product of the matrix $A$ by its complex conjugate and
\[
    DR(x) = \frac{dR(x)}{dx} + [Q(x), R(x)].
\]

Let us now describe how the matrices $\sigma_{L}(x)$ and $\sigma_{R}(x)$ can be obtained in the first place. They are solutions of the differential equations
\begin{eqnarray}\label{DensityLindblad}
    &&\frac{d\sigma_{L}(x)}{dx} = Q^{\dagger}(x) \sigma_{L}(x) + {\rm H.c.}
    + R^{\dagger}(x)\sigma_{L}(x)R(x), 
    \\
    &&\frac{d\sigma_{R}(x)}{dx} = - Q(x)\sigma_{R}(x) 
    - {\rm H.c.}
    - R(x)\sigma_{R}(x)R^{\dagger}(x) 
\end{eqnarray}
with the boundary conditions $\sigma_{L}(0) = |\nu_{L}\rangle \langle \nu_{L} |$ and $\sigma_{R}(L) = |\nu_{R}\rangle \langle \nu_{R} |$.
In the following, we call these equations (and their analogs for other density matrices) as the Lindblad equations, since under certain gauges they reduce to the Lindblad master equation. In the numerical optimization, we integrate these equations approximately using the scheme from Ref.~\cite{PhysRevLett.128.020501}, but we can also obtain the exact solution, which we employ below in the derivation of the energy expectation value. 

By introducing the matrix 
\begin{equation}\label{eq:Tmatrix}
    T(u) = Q(u) \otimes 1 + 1 \otimes \overline{Q}(u)+ R(u) \otimes \overline{R}(u),
\end{equation}
we can write the density matrices in a compact form:
\begin{eqnarray}
    \sigma_{L}(x) 
    &=& \sigma_{L}(0) P\exp \int_{0}^{x} T(u)du,
    \label{eq:sigmaL}
    \\
    \sigma_{R}(x) 
    &=& P\exp \left[\int_{x}^{L} T(u) du\right] \sigma_{R}(L).
    \label{eq:sigmaR}
\end{eqnarray}
For the computation purpose, we can now express the energy expectation value as
\begin{multline}\label{Energy}
    \langle E \rangle = w\int_{0}^{L}dx {\langle \sigma_{L}(x)|H(x)|\sigma_{R}(x)\rangle}
    \\
    + w\int_{0}^{L} dx \int_{x}^{L} dy  U(y,x) 
    \langle \sigma_{L}(x)|R(x)\otimes \overline{R(x)}  
    \\ 
    \times
    P\exp
    \left[\int_{x}^{y} du T(u)\right]
    R(y)\otimes \overline{R(y)}
    |\sigma_{R}(y)\rangle,
\end{multline}
where $w\equiv1/\langle \sigma_{L}(x)|\sigma_{R}(x)\rangle=1/\langle Q,R| Q,R \rangle$ is the wave-function normalization factor. Its independence on the coordinate $x$ can be verified directly from the inner product of the wave functions expressed by Eqs.~\eqref{eq:sigmaL} and \eqref{eq:sigmaR}.
The first integral in Eq.~\eqref{Energy} corresponds to expectation value of the local part of the Hamiltonian operator, where $H(x)$ is defined from local part of Eq.~\eqref{Hamiltonian} according to the rules~\eqref{Rules} as follows:
\begin{multline}
    H(x) = \frac{1}{2} DR(x) \otimes \overline{DR(x)}
    \\ 
    + (V(x)-\mu)R(x) \otimes \overline{R(x)} + g R(x)^{2} \otimes \overline{R(x)}^{2}
\end{multline}
The second integral in Eq.~\eqref{Energy} corresponds to the non-local long-range interaction. 

The next step is to represent the energy as a sum of scalar products of local quantities (which are described by some kind of differential equations).
To this end, we choose a certain point $z$ and divide the energy into three parts: 
(i) expectation values of the operators determined solely to the left from the point $z$, 
(ii) expectation values of operators determined solely to the right from the point $z$, and (iii) operators acting on both sides from the point $z$ (the last part naturally appears in the computation of the expectation value of non-local long-range interactions). 
The first part can be written as
\begin{multline}\label{LeftEnergy}
    w\int_{0}^{z} \langle \sigma_{L}(x)|H(x)|\sigma_{R}(x)\rangle dx
    \\
    + w \int_{0}^{z}dy \int_{0}^{y} dx U(y,x) \langle \sigma_{L}(x)|R(x)\otimes \overline{R(x)}
    \\
    \times P\exp{\left[\int_{x}^{y} T(u) du\right]}   R(y)\otimes \overline{R(y)}|\sigma_{R}(y)\rangle.
\end{multline}
The second part has a similar form,
\begin{multline}\label{RightEnergy}
    w\int_{z}^{L}
    {\langle \sigma_{L}(x)|H(x)|\sigma_{R}(x)\rangle}
    dx
    \\
    + w\int_{z}^{L} dx \int_{x}^{L} dy U(y,x) \langle \sigma_{L}(x)|R(x)\otimes \overline{R(x)}
    \\
    \times P\exp{\left[\int_{x}^{y} T(u) du\right]}   R(y)\otimes \overline{R(y)}|\sigma_{R}(y)\rangle
    .
\end{multline}
And the third part reads as
\begin{multline}\label{LongEnergy}
    w\int_{0}^{z} dx \int_{z}^{L} dy U(y,x) \langle \sigma_{L}(x)|R(x)\otimes \overline{R(x)} 
    \\
    \times P\exp{\left[\int_{x}^{z} T(u) du\right]}
    P\exp{\left[\int_{z}^{y} T(v) dv\right]}   
    \\ 
    \times R(y)\otimes \overline{R(y)}|\sigma_{R}(y)\rangle
    .
\end{multline}

The first part of Eq.~\eqref{LeftEnergy} can be represented as $w\langle H_{L}(z)| \sigma_{R}(z)\rangle$, where the matrix $H_{L}(z)$ is defined according the following expression:
\begin{multline}\label{HL}
    \langle H_{L}(z)| =     \int_{0}^{z} \langle \sigma_{L}(x)|H(x)
    P\exp{\left[\int_{x}^{z} T(u) du\right]} dx
    \\
    + \int_{0}^{z}dy \int_{0}^{y}dx   U(y,x) \langle \sigma_{L}(x)|R(x)\otimes \overline{R(x)} 
    \\
    \times P\exp{\left[\int_{x}^{y} T(u) du\right]}
    R(y)\otimes \overline{R(y)} 
    \\ 
    \times P\exp{\left[\int_{y}^{z} T(u) du\right]}
    .
\end{multline}

The second part of Eq.~\eqref{RightEnergy} can be written similarly as $w\langle \sigma_{L}(z)| H_{R}(z)\rangle$. The third part~\eqref{LongEnergy}, in general, can not be cast in the form $\sum_{i} \langle U_{L,i}(z)|U_{R,i}(z)\rangle$, since the interaction potential $U(x,y)$ connects the left and right parts together. 
But in case of the factorizable potential, $U(x,y) = \sum_{i} f_{i}(x)g_{i}(y)$, the third part~\eqref{LongEnergy} can be represented in the form $\sum_{i} \langle U_{L,i}(z)|U_{R,i}(z)\rangle$ with $\langle U_{L,i}(x)|$ and $|U_{R,i}(x)\rangle$ defined as follows:
\begin{eqnarray}\label{UL}
    \langle U_{L,i}(z)| &=& \int_{0}^{z} f_{i}(x) \langle \sigma_{L}(x)|R(x)\otimes \overline{R(x)} \nonumber
    \\
    &&\times
     P\exp{\left[\int_{x}^{z} T(u) du\right]}dx,
    \\
    \label{UR}
     |U_{R,i}(z)\rangle &=&    \int_{z}^{L} g_{i}(y)
      P\exp{\left[\int_{z}^{y} T(v) dv\right]}\nonumber
     \\ 
     &&\times R(y)\otimes \overline{R(y)}|\sigma_{R}(y)\rangle dy.
\end{eqnarray}

Let us now show that $H_{L}(x)$, $U_{L,i}(x)$, and $\sigma_{L}(x)$ form together a system of linear differential equations, which can be used to compute these in the same way, as $\sigma_{L}(x)$ was computed by using Eq.~\eqref{DensityLindblad}. 
For the derivation of equations, we can simply differentiate Eqs.~\eqref{HL} and \eqref{UL}. This yields
 \begin{multline}\label{ULindblad}
     \frac{dU_{L,i}(z)}{dz}
     = Q^{\dagger}(z) U_{L,i}(z) + U_{L,i}(z) Q(z) 
     \\
     + R^{\dagger}(z) U_{L,i}(z) R(z)
     + f_{i}(z) R^{\dagger}(z) \sigma_{L}(z) R(z).
 \end{multline}
And for the energy $H_{L}(z)$ we obtain the equation
\begin{multline}\label{HLindblad}
    \frac{dH_{L}(z)}{dz} = Q^{\dagger}(z) H_{L}(z) + H_{L}(z) Q(z) + R^{\dagger}(z) H_{L}(z) R(z)
    \\
    + \sigma_{L}(z) H(z)+\sum_{i} g_{i}(z) R^{\dagger}(z) U_{L,i}(z) R(z).
\end{multline}
These equations must be supplemented with the boundary conditions $U_{L,i}(0) = 0$ and $H_{L}(0)= 0$. As for $U_{R,i}(z)$ and $H_{R}(z)$, we obtain completely analogous equations with opposite signs and with the interchanged role of $f_{i}(z)$ and $g_{i}(z)$. Using these equations and discretization scheme proposed in Ref.~\cite{PhysRevLett.128.020501}, we then compute the energy expectation value. 

Interaction potentials of the form $U(x,y) = \sum_{i} f_{i}(x)g_{i}(y)$ can appear in certain many-mode cavity systems~\cite{Mottl2014RotontypeMS, Gopalakrishnan_2009}. Still, according to the most of physical applications, we are interested in the class of potentials, which depend only on the relative position of two particles, $U(x,y) = U(y-x)$. 
To represent this potential in the factorized form, we can approximate $U(y-x)$ by the sum of exponents (see also Ref.~\cite{PhysRevB.78.035116} in the lattice context), 
\begin{equation}\label{approximation}
    U(y-x) \approx \sum_{i=1}^{n} A_{i} 
    \exp{\left[-a_{i}(y-x)\right]}.
\end{equation}
This approximation is explicitly factorizable, but the functions $f_{i}(x) = \exp(a_{i}x)$ and $g_{i}(y) = \exp(-a_{i}y)$ are problematic, since one of them can quickly become exponentially small, while another one becomes exponentially large. 
We can solve this problem by expressing $\exp{[-a_{i}(y-x)]} = \exp{[-a_{i}(y-z)]} \times \exp{[-a_{i}(z-x)]}$. 
Relying on this decomposition, we redefine the matrices $U_{L,i}(z)$ and $U_{R,i}(z)$,
\begin{multline}
    \langle U_{L,i}(z)| = \int_{0}^{z} \langle \sigma_{L}(x)|R(x)\otimes \overline{R(x)} \exp{[-a_{i}(z-x)]}
    \\
    \times  P\exp{\left[\int_{x}^{z} T(u) du\right]}
    dx,
\end{multline}
\begin{multline}
         |U_{R,i}(z)\rangle =    \int_{z}^{L} \exp{[-a_{i}(y-z)]} 
          P\exp{\left[\int_{z}^{y} T(v) dv\right]}
         \\ 
         \times R(y)\otimes \overline{R(y)}|\sigma_{R}(y)\rangle dy.
\end{multline}
The new matrices $U_{L,i}(z)$ obey the linear differential equations
 \begin{multline}\label{ULindbladCorrected}
     \frac{dU_{L,i}(z)}{dz} = Q^{\dagger}(z) U_{L,i}(z) + U_{L,i}(z) Q(z) 
     \\
      + R^{\dagger}(z) U_{L,i}(z) R(z) + R^{\dagger}(z) \sigma_{L}(z) R(z) - a_{i} U_{L,i}(z).
 \end{multline}
Note that equations for different $i$ are completely independent, thus can be solved in parallel. 

With the new $U_{L,i}(z)$, Eq.~\eqref{HLindblad} changes to the following form:
\begin{multline}\label{EnergyLindblad}
    \frac{dH_{L}(z)}{dz} = Q^{\dagger}(z) H_{L}(z) + H_{L}(z) Q(z) + R^{\dagger}(z) H_{L}(z) R(z)
    \\
    + \sigma_{L}(z) H(z)+\sum_{i} A_{i} R^{\dagger}(z) U_{L,i}(z) R(z).
\end{multline}
Using the introduced notations, we can express the energy determined by Eq.~\eqref{Energy} as follows:
\begin{eqnarray}\label{EnergyExpectation}
    \langle E \rangle =
    w\left[
        \langle H_{L}(z)|\sigma_{R}(z)\rangle + \langle \sigma_{L}(z)|H_{R}(z)\rangle
    \right]\nonumber
    \\
    + w\sum_{i=1}^{n} A_{i} \langle U_{L,i}(z)|U_{R,i}(z)\rangle.
\end{eqnarray}
Note that Eq.~\eqref{EnergyExpectation} is independent of $z$ due to the Lindblad equations for all matrices involved in the expression. 

The next goal is to compute energy gradients. We delegate the explicit derivation of gradients to Appendix~\ref{AppA} due to complexity of the corresponding expressions. At the same time, we note here that these gradients can be expressed in terms of certain integrals of the  matrices $H_{L}(x)$, $H_{R}(x)$, $U_{L,i}(x)$, $U_{R,i}(x)$, $\sigma_{L}(x)$, and $\sigma_{R}(x)$.
Let us also note that in contrast to the available possibilities to employ the numerical automatic differentiation packages, the explicit expressions in Appendix~\ref{AppA} enable more efficient parallelization and independent control of the accuracy of gradients and energy calculations.

Hence, the obtained equations allow us to compute the energy expectation value for both the translationally-invariant interaction potentials and the cavity-like interactions.
It is clear now that one can perform calculations with the same interactions, which are tractable with the matrix-product operators (MPO) in the lattice context. We discuss this analogy in more detail in Appendix~\ref{AppB}.

\section{Model}
To benchmark the developed approach, we choose the Calogero-Moser rational model~\cite{calogero1969ground, calogero1969solution,calogero1971solution,calogero1975exactly,sutherland1971quantum, Polychronakos1992ExchangeOF} (see also the review~\cite{Polychronakos_2006}). The corresponding Hamiltonian for the system of $N$ interacting particles reads as
\begin{equation}\label{eq:Hcm}
    H_{\rm CM} = -\frac{1}{2}\sum_{i=1}^{N}\frac{\partial^{2}}{\partial x_{i} ^{2}} +
    \sum_{i=1}^{N} \frac{\omega^2 x_{i}^{2}}{2} +\sum_{1 \leq i < j \leq N} \frac{ l(l-1)}{(x_{i}-x_{j})^{2}}.
\end{equation}
This model includes the external harmonic confinement and the two-body long-range interactions with the singular potential $U(x,y) = \frac{l(l-1)}{(y-x)^{2}}$.
We restrict ourselves to the repulsive case with $l>1$. 
This model is exactly solvable and its ground-state energy $E_{N}$ is given by
\begin{equation}\label{ExactEnergy}
    E_{N} = \omega \left[\frac{N}{2} + l \frac{N(N-1)}{2}\right].
\end{equation}
The bosonic ground-state wave function $\psi_{0}$ can be expressed in the Jastrow form:
\begin{equation}\label{ExactWF}
    \psi_{0} = \prod_{1\leq i < j \leq N} |x_{i}-x_{j}|^{l} \prod_{i=1}^{N} \exp{(-\omega x_{i}^{2}/2)}.
\end{equation}
This wave function vanishes exponentially in the limit of large $x_{i}$ due to the external trapping potential and decreases polynomially for two particles approaching each other.
Note that in the limit $l \to 1$, the wave function turns into the wave function of free fermions, since strong repulsive interactions enforce vanishing of the wave function for two coinciding particles. 

Due to the exponential decrease of the wave function at large $x_{i}$, we can restrict the system to the finite interval~$[0,L]$ with a corresponding shift of the minimum of harmonic potential to its center, $V(x) = \omega^{2}(x-L/2)^{2} /2$. 
If $L$ and $\omega$ are sufficiently large,
at the boundaries $x_{i} = \{0, L\}$ we can enforce the vanishing wave-function boundary conditions, which do not change the energy and wave-function behavior.

The main difficulty with the application of the cMPS ansatz to the Calogero-Moser model is the singularity $1/x^{2}$ of the interaction potential  at small $x$. 
Following the discussed procedure [see Eq.~\eqref{approximation}], we need to approximate this potential by a sum of exponents, $1/x^{2} \approx \sum_{i}^n A_{i} \exp{[-a_{i}x]}$. 
This approximation can not hold in the vicinity of $x = 0$, thus one can reliably approximate the potential only at $x>\varepsilon$ with a certain small~$\varepsilon$. 
Therefore, in our analysis  we approximate the potential on the interval $[\varepsilon, L]$. This approximation fixes the parameters $a_{i}$ and $A_{i}$. After that, in all calculations we use the approximate potential defined as a sum of exponents on the whole interval $[0,L]$.

We can use the finite sum of exponents as a definition of the approximate interaction potential for all $x$.  However, in the vicinity of $x = 0$ the approximation results in a large but finite value of the potential, in contrast to the singular behavior of the real interaction potential.
Still, we can argue that the error in the given approximation at small $x$ insignificantly impacts on the variational energy and wave function.  We can expect deviations between the exact wave function and the variational cMPS only at $|x_{i} - x_{j}| \lesssim \varepsilon$. In this regime,
the exact wave function vanishes polynomially as $\psi_{0} \propto |x_{i} - x_{j}|^{l}$. 
We also established numerically that the cMPS wave function vanishes to high accuracy in the presence of large but finite potential core of the radius~$\varepsilon$. 
From this we can argue that the exact and variational wave functions deviate from each other in the region of very small densities leading to negligibly small absolute errors. 

In the next section,  we investigate convergence of the energy with respect to $\varepsilon$ and the cut-off number $n$ of exponents in the approximation in more detail.

\section{Results and benchmarks}

Our approach to long-range interacting systems is not exactly variational, since it depends crucially on the approximation of the interaction potential with a sum of exponents. 
This approximation can underestimate the exact interaction (e.g., near the core of the singular potential), thus the energy obtained within the numerical procedure can be smaller than the true ground-state energy. Certainly, the energy of the variationally obtained cMPS computed with the exact Hamiltonian is always larger than the energy of the exact ground state, but the energy of cMPS computed with the approximate Hamiltonian can be smaller. 
Here, we discuss the energy $E_{num}$, which is computed with the cMPS and approximate Hamiltonian (since the computation with the cMPS and exact Hamiltonian is difficult due to the singular potential). 
This numerical energy $E_{num}$ is then compared to the exact analytical results given by Eq.~\eqref{ExactEnergy}. 

There are several sources of errors we would like to point out. 
The first one originates from a finite number of variational parameters, which can be too small to represent all peculiarities of the exact ground state. 
The number of variational parameters is controlled by the bond dimension $D$ and by the grid size. 
The second source of errors is the restriction of the system to the finite spatial interval (while exact solutions correspond to the infinite system in a certain trapping potential). 
This truncation can be justified {\it aposteriori}, if the particle density obtained from the optimized cMPS vanishes exponentially at the boundary.
The third source of errors is the approximation of the interaction potential. In case of a singular potential, this approximation depends both on the core size $\varepsilon$ and on the number $n$ of exponents in Eq.~\eqref{approximation}. 
Note that the dependence on these parameters highly varies for different potentials and depends on the approximation method or on the metrics to estimate the reliability of the approximation. 
In particular, one can not rely solely on the maximal deviation between the exact and approximate potentials.
On the one hand, in the vicinity of a singular core, these deviations are always extremely large, but probabilities of finding particles on these interparticle distances are very small. 
On the other hand, even small deviations of potentials on  moderate interparticle distances can lead to noticeable errors. Since average interparticle distances depend on the trapping potential and interaction strength, an accurate approximation of the potential for a particular set of parameters may become not optimal for another one. 

\begin{figure}[t]
  \includegraphics[width=\linewidth]{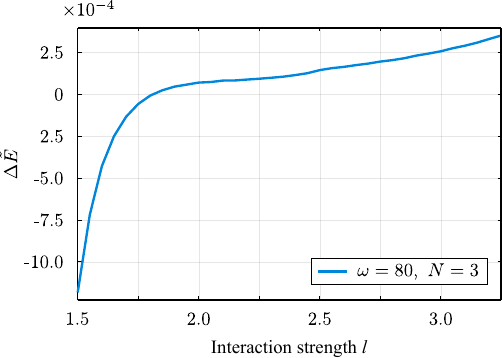}  
  \caption{\label{fig:EnergyError}%
    Relative difference $\Delta \tilde{E}=(E_{num}-E_N)/E_N$ between the computed cMPS energy~$E_{num}$ and the analytical prediction~\eqref{ExactEnergy} as a function of the interaction strength~$l$. Other parameters are $\omega = 80$, $N=3$, $L=1$, $D=12$, $N_{mesh} = 170$, $n=8$, and $\varepsilon = 0.025$.
    }
\end{figure}

First, let us determine whether the developed approach to describe long-range interacting systems works in principle, investigate the sources of errors, and analyze which of them are the most influential for different values of the model parameters.
We start with a small number of particles, $N=3$ for $\omega = 80$ and $L = 1$. 
We also fix $D=12$, $N_{mesh} = 170$, $n=8$, and $\varepsilon = 0.025$. 
For these parameters the relative difference $\Delta \tilde{E}=(E_{num}-E_N)/E_N$ between the computed cMPS energy~$E_{num}$ and the exact result~\eqref{ExactEnergy} is shown in Fig.~\ref{fig:EnergyError}. 
Note that there is an additional possible source of error due to not exactly integer number of particles in the cMPS wave function. The particle number is regulated by an adjustment of the chemical potential $\mu$ in the auxiliary term $\mu N$ added to the Hamiltonian~\eqref{eq:Hcm}. 
In the performed calculations with $N=3$, the absolute error in the number of particles is typically about $10^{-6}$. 
This deviation in the number of particles introduces a relative error in the energy, which is one order of magnitude smaller than errors from other sources. 
We observe that the relative energy difference is generally rather small confirming that the developed approach is sufficiently accurate. 

At small interaction strengths $l$, the relative error is negative and grows rapidly. 
This is a sign of underestimation of the repulsive interaction potential. 
At small interaction strength, two interacting particles are able to reach relative distances smaller than $\varepsilon$, thus numerical difference between the exact and approximate potentials in this region causes errors in energies and wave functions. 
This error can be mitigated by decreasing $\varepsilon$. 
At large $l$, the relative error is positive and grows slowly. 
The main reason for this growth is the truncation of the wave function at the boundaries. In particular, in the region $l\gtrsim3.5$ particles experience a strong repulsion, which becomes insufficiently compensated by the trapping potential (with the given amplitude $\omega = 80$) in order to completely suppress the wave function at the boundaries.

\begin{figure}[t]
  \includegraphics[width=\linewidth]{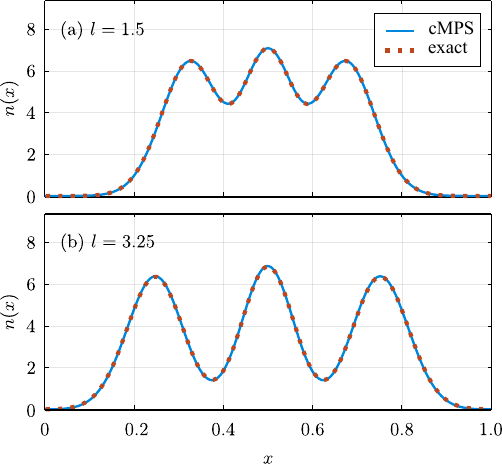}
  \caption{\label{fig:Density1.5}%
    Particle density distributions obtained by the cMPS numerical approach and analytic form of the wave function~\eqref{ExactWF} at $l=1.5$ (a) and $l=3.25$ (b). Other parameters are the same as in Fig.~\ref{fig:EnergyError}. 
    }
\end{figure}

In Fig.~\ref{fig:Density1.5}
we show the density distributions at small and large interactions.
For comparison, we also plot the density distributions obtained from the numerical integration of the exact wave function~\eqref{ExactWF}, which visually coincide with the ones from the cMPS approach.
At small interaction [see Fig.~\subfigref{fig:Density1.5}{a}], the wave functions are exponentially suppressed at the boundaries, while at large interaction [see Fig.~\subfigref{fig:Density1.5}{b}] the density vanishes significantly slower at the edges. This is the reason for the deviations in the ground-state energy at large $l$ (see Fig.~\ref{fig:EnergyError}).
At the same time, the density distribution in Fig.~\subfigref{fig:Density1.5}{b} demonstrates more pronounced minima than in Fig.~\subfigref{fig:Density1.5}{a}. 
It means that the average interparticle distances are rather large, thus specific details of the parametrization of the core of the singular potential become less important.

\begin{figure}[t]
  \includegraphics[width=\linewidth]{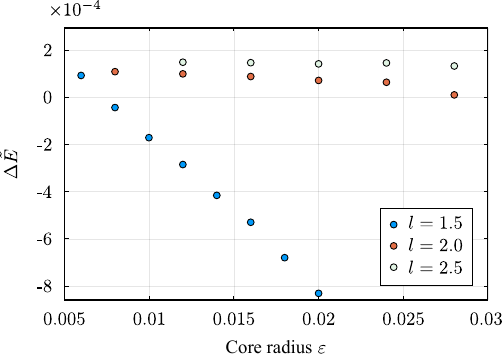}
  \caption{\label{fig:RadiusInfluence}%
    Relative difference $\Delta \tilde{E}=(E_{num}-E_N)/E_N$ between the computed cMPS energy~$E_{num}$ and the analytical prediction~\eqref{ExactEnergy} as a function of the potential core radius~$\varepsilon$  at three different interaction strengths~$l$. Other parameters are the same as in Fig.~\ref{fig:EnergyError}.
    }
\end{figure}

\begin{figure}[t]
  \includegraphics[width=\linewidth]{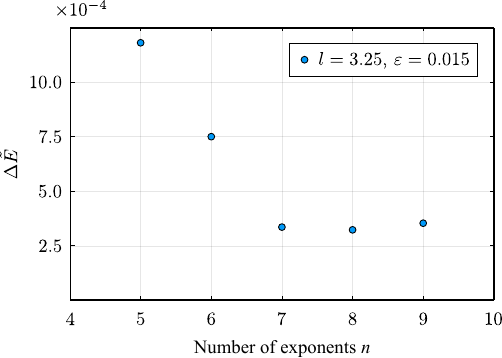}
  \caption{\label{fig:NexpInfluence}%
    Relative difference between numerically computed energy of optimized cMPS $E_{num}$ and analytical prediction for different numbers of exponents $n$ in the approximation of the potential at $l = 3.25$ and $\varepsilon=0.015$. Other parameters are 
    the same as in Fig.~\ref{fig:EnergyError}.
    }
\end{figure}

To analyze the dependence of the variational energy on the radius~$\varepsilon$, we fix $l=1.5$, since at low $l$ the influence of the core is larger, and optimize the wave function for different $\varepsilon$. We show the corresponding dependence of the relative energy difference $\Delta \tilde{E}$ in Fig.~\ref{fig:RadiusInfluence}. 
From it we can conclude that the dependence on $\varepsilon$ is approximately linear.
In particular, at small $\varepsilon$ the error becomes positive, as it should be in the variational approaches. 
At large $l$, the influence of $\varepsilon$ is less pronounced (see also Fig.~\ref{fig:RadiusInfluence} for $l=2.0$ and $l=2.5$).
In this regime, we can also test the dependence of the relative error~$\Delta \tilde{E}$ on the number of exponents $n$ in the approximation of the potential with the fixed $\varepsilon$, which is given in Fig.~\ref{fig:NexpInfluence}.
It shows the dependence of the relative error on the number of exponents at $l=3.25$ and $\varepsilon=0.015$ (other parameters are the same as in the preceeding analysis). 
We also verified that that the energy does not vary strongly with the change of $\varepsilon$. 
The energy difference decreases with $n$ till it reaches approximately a constant value (which can further increase due to possible underestimation of the interaction strength at certain $x$). 
At smaller number of terms in the approximation ($n\leq5$), one must change $\varepsilon$ to larger values to obtain more accurate energies, since the approximation method involves all the variational freedom to approximate the potential core and contains large errors in the tail of the interaction potential.
Note that the results in Fig.~\ref{fig:NexpInfluence} should only be viewed as qualitative, since the numerical accuracy significantly depends on specific values of parameters $\varepsilon$, $N$, $l$, and $\omega$, as well as on the methodology of approximation.

\begin{figure}[t]
  \includegraphics[width=\linewidth]{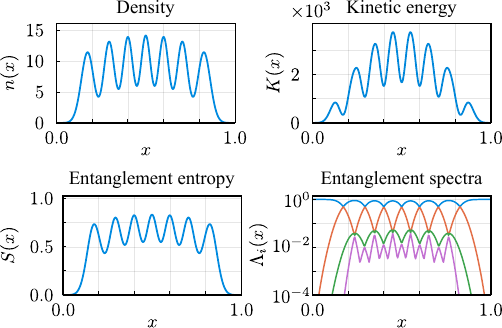}
  \caption{\label{fig:N=7}%
    Spatial distributions of physical characteristics of the system at
    $N=7$, $l = 2.7$, $\omega = 185$, $n=9$, $\varepsilon = 0.01$, $D = 30$, $N_{mesh} = 250$, and $L = 1$. In the entanglement spectra, only four largest eigenvalues are shown.}
\end{figure}

After we analyzed the accuracy of the developed approach in a relatively dilute regime ($N=3$), we can test it on systems with a larger number of particles. 
To this end, we take $N=7$, $\omega = 185$, $l = 2.7$ and determine the wave function variationally. 
With the increased bond dimension $D=30$ we obtain the relative energy error $\Delta\tilde{E} = 7 \times 10^{-5}$.
In Fig.~\ref{fig:N=7} we show several relevant physical characteristics of the system determined by this wave function: the particle density, the kinetic energy and the entanglement. 
We compute the entanglement spectra $\Lambda_i(x)$ using diagonalization of the matrix $\sigma_{L}(x) \sigma_{R}(x)$~\cite{PhysRevB.83.245134}. The entanglement entropy determined as $S(x)=-\sum_{i} \Lambda_{i}(x) \log{\Lambda_{i}(x)}$ generally mimics the density distribution, while the two largest eigenvalues $\Lambda_{1,2}$ cross at the density maxima. This is a general observation holding also for other model parameters of the system under study.

\section{Conclusions and Outlook}
In this study, we developed the methodology to apply the cMPS computational approach to inhomogeneous one-dimensional systems with long-range interactions. 
We established that the long-range interactions with a potential expressed in the form of a sum of exponents (or cavity-like interactions) can be efficiently simulated in the exact variational manner. From this fact we proposed an approximate general scheme for the many-body systems with the interaction potentials of an arbitrary form. 
The proposed methodology is also compared to the lattice DMRG studies of long-range interacting systems.

We benchmarked the numerical approach on the exactly solvable Calogero-Moser model in the external harmonic potential. This model contains a singular interaction potential between particles.
We outlined how the cMPS methodology can be applied to systems with similar singular interaction potentials and confirmed the validity and accuracy of the method on both the variational energy and the ground-state local observables such as the particle density.

There are several potential research directions we would like to pursue. The first one concerns an application of the method to studies of phase transitions or dualities in the systems of bosons in the cavity~\cite{CavityFermionization, PhysRevA.100.013611}.
One can also apply the methodology to ultracold dipolar \cite{Chomaz2022} and Rydberg \cite{Marcassa2014} gases in one-dimensional traps.

Within this study, we employed the global gradient optimization of the wave function. However, the similarity to the DMRG allows to at least partially generalize the local optimization with sweeps to the continuous case. This is another interesting direction for a thorough and separate analysis.

\begin{acknowledgments}
The authors acknowledge support from 
the National Research Foundation of Ukraine, Grant No.~0120U104963,
the Ministry of Education and Science of Ukraine, Research Grant No.~0122U001575, and the National Academy of Sciences of Ukraine, Project No. 0121U108722.
\end{acknowledgments}

\appendix
\section{Derivation of gradients}\label{AppA}
Let us discuss in more detail the procedure of calculating the energy gradients in terms of the variational parameters $R_{k}$ and $Q_{k}$ of the cMPS wave function. 
These parameters describe the wave function only on a small spatial interval $[x_{k-1}, x_{k+1}]$. 
The matrices $\sigma_{L}(x)$, $H_{L}(x)$, and $U_{L}(x)$ are described by the Lindblad equations in a conventional form.
It means that these matrices are independent of $R_{k}$ and $Q_{k}$ for $x \in [0, x_{k-1}]$. 
The same is valid for the matrices $\sigma_{R}(x)$, $H_{R}(x)$, and $U_{R}(x)$, since they are independent of $R_{k}$ and $Q_{k}$ for $x \in [x_{k+1},L]$. 

To obtain the gradients, we express the energy $E$ [see Eq.~\eqref{EnergyExpectation}] using only the matrices defined at $x=x_{k}$:
\begin{widetext}
 \begin{equation}\label{EnergyForGradient}
    E = w\left[
    {\langle H_{L}(x_{k})| \sigma_{R}(x_{k})\rangle + \langle \sigma_{L}(x_{k})| H_{R}(x_{k})\rangle + \sum_{i} A_{i} \langle U_{L,i}(x_{k})| U_{R,i}(x_{k})\rangle}
    \right]
    .
\end{equation}
The next step is to calculate the derivatives of the type $\langle \nabla_{R_{k}} \sigma_{L}(x_{k})| H_{R}(x_{k})\rangle$. 
To this end, we express the Lindblad equation~\eqref{DensityLindblad} for the density matrix $\sigma_{L}(x)$ in the finite difference form with the step $\Delta x$,
\begin{eqnarray}
    \sigma_{L}(x_{k}) = \sigma_{L}(x_{k} - \Delta x) + Q(x_{k} - \Delta x)^{\dagger} \sigma_{L}(x_{k} - \Delta x) + \sigma_{L}(x_{k} - \Delta x)Q(x_{k}-\Delta x)
    \nonumber
    \\ +  R(x_{k}-\Delta x)^{\dagger} \sigma_{L}(x_{k} - \Delta x) R(x_{k}-\Delta x).
    \label{FiniteDifference}
\end{eqnarray}
We can now take the derivative of Eq.~\eqref{FiniteDifference} by ${R_{k}}$ and use the compact notation~\eqref{eq:Tmatrix},
\begin{multline}\label{Derivative}
    \nabla_{R_{k}} \sigma_{L}(x_{k}) = \nabla_{R_{k}} \sigma_{L}(x_{k} - \Delta x)  + Q(x_{k} - \Delta x)^{\dagger} \nabla_{R_{k}} \sigma_{L}(x_{k} - \Delta x)  + \nabla_{R_{k}} \sigma_{L}(x_{k} - \Delta x)Q(x_{k}-\Delta x) \\ 
    + R(x_{k}-\Delta x)^{\dagger} \nabla_{R_{k}} \sigma_{L}(x_{k} - \Delta x) R(x_{k}-\Delta x) 
    + R(x_{k}-\Delta x)^{\dagger}  \sigma_{L}(x_{k} - \Delta x) \nabla_{R_{k}} R(x_{k}-\Delta x) \\ 
    = \nabla_{R_{k}} \sigma_{L}(x_{k} - \Delta x) \exp{[T(x_{k}) \Delta x]}  + R(x_{k}-\Delta x)^{\dagger}  \sigma_{L}(x_{k} - \Delta x) \nabla_{R_{k}} R(x_{k}-\Delta x)
\end{multline}

Next, it is necessary to calculate $\nabla_{R_{k}} \sigma_{L}(x_{k} - \Delta x)$, but this can be performed by using the same finite difference formula~\eqref{FiniteDifference}. 
By repeating the procedure $n$ times, we obtain the following equation:
\begin{multline}\label{Der2}
    \nabla_{R_{k}} \sigma_{L}(x_{k}) = \nabla_{R_{k}} \sigma_{L}(x_{k} - n\Delta x)  \prod_{i=n}^{1} \exp{[T(x_{k} -i \Delta x) \Delta x}] 
    \\
    + \sum_{i=1}^{n} R(x_{k}-i \Delta x)^{\dagger}  \sigma_{L}(x_{k} - i \Delta x) \nabla_{R_{k}} R(x_{k}-i \Delta x)   \prod_{j=i}^{1} \exp{[T(x_{k} -j \Delta x) \Delta x]}.
\end{multline}

If $n\Delta x > |x_{k} - x_{k-1}|$, then $\nabla_{R_{k}} \sigma_{L}(x_{k} - n\Delta x) = 0$ due the conventional structure of the Lindblad equation and only the second term remains in Eq.~\eqref{Der2}. In the continuous limit $\Delta x \to 0$, $n\Delta x = |x_{k} - x_{k-1}|$ the sum in Eq.~\eqref{Der2} transforms into the integral. Therefore, we finally obtain the closed expression for $\nabla_{R_{k}} \sigma_{L}(x_{k})$,
\begin{equation}
    \nabla_{R_{k}} \sigma_{L}(x_{k}) = \int_{x_{k-1}}^{x_{k}} dx R(x)^{\dagger} \sigma_{L}(x) \nabla_{R_{k}} R(x) P\exp{\left(\int_{x}^{x_{k}} T(u)du \right)} .
\end{equation}
The same procedure leads us to the simple expression for $\nabla_{Q_{k}} \sigma_{L}(x_{k})$,
\begin{equation}
    \nabla_{Q_{k}} \sigma_{L}(x_{k}) = \int_{x_{k-1}}^{x_{k}} dx \sigma_{L}(x) \nabla_{Q_{k}} Q(x)  P\exp{\left(\int_{x}^{x_{k}} T(u) du\right)} .
\end{equation}

The next step is to compute the derivatives of the type $\langle \nabla_{R_{k}} U_{L}(x_{k})| U_{R}(x_{k})\rangle$. 
Using the finite difference approximation to Eq.~\eqref{ULindbladCorrected} and the derivatives of $\sigma_{L}(x)$ obtained above, we arrive at
\begin{multline}\label{DerUR}
    \nabla_{R_{k}} U_{L,i}(x_{k})  =  \int_{x_{k-1}}^{x_{k}} dx R(x)^{\dagger} U_{L,i}(x) \nabla_{R_{k}} R(x) P\exp{\left[\int_{x}^{x_{k}}(T(z)-a_{i})dz\right]}
    \\ 
    + \int_{x_{k-1}}^{x_{k}} dx R(x)^{\dagger} \sigma_{L}(x) \nabla_{R_{k}} R(x) P\exp{\left[\int_{x}^{x_{k}}(T(z)-a_{i})
    dz\right]}
    \\
    + \int_{x_{k-1}}^{x_{k}} dx \int_{x_{k-1}}^{x} dy R(y)^{\dagger} \sigma_{L}(y) \nabla_{R_{k}} R(y) P\exp{\left[\int_{y}^{x}T(u)du\right]} R(x) \otimes \overline{R(x)} P\exp{\left[\int_{x}^{x_{k}}(T(z)-a_{i})
    dz\right]},
\end{multline}
\begin{multline}\label{DerUQ}
    \nabla_{Q_{k}} U_{L,i}(x_{k})  =  \int_{x_{k-1}}^{x_{k}} dx  U_{L,i}(x) \nabla_{Q_{k}} Q(x) P\exp{\left[\int_{x}^{x_{k}}(T(z)-a_{i})dz\right]}
    \\ 
    + \int_{x_{k-1}}^{x_{k}} dx \int_{x_{k-1}}^{x} dy  \sigma_{L}(y) \nabla_{Q_{k}} Q(y) P\exp{\left[\int_{y}^{x}T(u)du\right]} R(x) \otimes \overline{R(x)} P\exp{\left[\int_{x}^{x_{k}}(T(z)-a_{i})
    dz\right]} .
\end{multline}
Let us express the derivatives of $H_{L}(x_{k})$,
\begin{multline}\label{DerHR}
    \nabla_{R_{k}} H_{L}(x_{k})  = \int_{x_{k-1}}^{x_{k}} dx R(x)^{\dagger} H_{L}(x) \nabla_{R_{k}} R(x) P\exp{\left[\int_{x}^{x_{k}} T(z)dz\right]} 
    \\
    + \sum_{i=1}^{n} A_{i}\int_{x_{k-1}}^{x_{k}} dx R(x)^{\dagger} U_{L,i}(x) \nabla_{R_{k}} R(x) P\exp{\left[\int_{x}^{x_{k}}T(z)
    dz\right]} 
    + \int_{x_{k-1}}^{x_{k}} dx \sigma_{L}(x) \nabla_{R_{k}} H(x) P\exp{\left[\int_{x}^{x_{k}}T(z)
    dz\right]}
    \\ 
    + \int_{x_{k-1}}^{x_{k}} dx \int_{x_{k-1}}^{x} dy R(y)^{\dagger} \sigma_{L}(y) \nabla_{R_{k}} R(y) P\exp{\left[\int_{y}^{x}T(u)du\right]}   H(x) P\exp{\left[\int_{x}^{x_{k}}T(z)
    dz\right]} + \\ + \sum_{i=1}^{n} A_{i} \int_{x_{k-1}}^{x_{k}} dx \int_{x_{k-1}}^{x} dy R(y)^{\dagger} U_{L,i}(y) \nabla_{R_{k}} R(y)  P\exp{\left[\int_{y}^{x}(T(u)-a_{i}) du\right]}  R(x) \otimes \overline{R(x)} P\exp{\left[\int_{x}^{x_{k}}T(z)
    dz\right]}
    \\ 
    + \sum_{i=1}^{n} A_{i} \int_{x_{k-1}}^{x_{k}} dx \int_{x_{k-1}}^{x} dy R(y)^{\dagger} \sigma_{L}(y) \nabla_{R_{k}} R(y)  P\exp{\left[\int_{y}^{x}(T(u)-a_{i}) du\right]}  R(x) \otimes \overline{R(x)} P\exp{\left[\int_{x}^{x_{k}}T(z)
    dz\right]} 
    \\
    + \sum_{i=1}^{n} A_{i} \int_{x_{k-1}}^{x_{k}} dx \int_{x_{k-1}}^{x} dy \int_{x_{k-1}}^{y} dt R(t)^{\dagger} \sigma_{L}(t) \nabla_{R_{k}} R(t) \times
    \\ 
    \times P\exp{\left[\int_{t}^{y} T(v)
    dv\right]}  R(y) \otimes \overline{R(y)} P\exp{\left[\int_{y}^{x}(T(u)-a_{i}) du\right]}   R(x) \otimes \overline{R(x)} P\exp{\left[\int_{x}^{x_{k}}T(z)
    dz\right]}.
\end{multline}
\begin{multline}\label{DerHQ}
    \nabla_{Q_{k}} H_{L}(x_{k})  = \int_{x_{k-1}}^{x_{k}} dx  H_{L}(x) \nabla_{Q_{k}} Q(x) P\exp{\left[\int_{x}^{x_{k}} T(z)dz\right]}  + \int_{x_{k-1}}^{x_{k}} dx \sigma_{L}(x) \nabla_{Q_{k}} H(x) P\exp{\left[\int_{x}^{x_{k}}T(z)
    dz\right]} + \\ 
    + \int_{x_{k-1}}^{x_{k}} dx \int_{x_{k-1}}^{x} dy  \sigma_{L}(y) \nabla_{Q_{k}} Q(y) P\exp{\left[\int_{y}^{x}T(u)du\right]}   H(x) P\exp{\left[\int_{x}^{x_{k}}T(z)
    dz\right]} + \\ + \sum_{i=1}^{n} A_{i} \int_{x_{k-1}}^{x_{k}} dx \int_{x_{k-1}}^{x} dy  U_{L,i}(y) \nabla_{Q_{k}} Q(y)  P\exp{\left[\int_{y}^{x}(T(u)-a_{i}) du\right]}  R(x) \otimes \overline{R(x)} P\exp{\left[\int_{x}^{x_{k}}T(z)
    dz\right]} + \\  + \sum_{i=1}^{n} A_{i} \int_{x_{k-1}}^{x_{k}} dx \int_{x_{k-1}}^{x} dy \int_{x_{k-1}}^{y} dt \sigma_{L}(t) \nabla_{Q_{k}} Q(t) \times \\ \times Pexp{\left[\int_{t}^{y} T(v)
    dv\right]}  R(y) \otimes \overline{R(y)} P\exp{\left[\int_{y}^{x}(T(u)-a_{i}) du\right]}   R(x) \otimes \overline{R(x)} P\exp{\left[\int_{x}^{x_{k}}T(z)
    dz\right]}
\end{multline}

Now, we can determine the sum 
  \begin{multline}\label{DerEQ}
     \langle \nabla_{Q_{k}} H_{L}(x_{k})| \sigma_{R}(x_{k})\rangle + \langle \nabla_{Q_{k}} \sigma_{L}(x_{k})| H_{R}(x_{k})\rangle  + \sum_{i} A_{i} \langle \nabla_{Q_{k}} U_{L,i}(x_{k})| U_{R,i}(x_{k})\rangle 
     \\ 
     = \int_{x_{k-1}}^{x_{k}} dx \langle  H_{L}(x)|\nabla_{Q_{k}} Q(x)  P\exp{\left[\int_{x}^{x_{k}} T(z)dz\right]}|\sigma_{R}(x_{k})\rangle 
     \\
     + \sum_{i=1}^{n} A_{i} \int_{x_{k-1}}^{x_{k}} dx  \langle U_{L,i}(x)|\nabla_{Q_{k}} Q(x) P\exp{\left[\int_{x}^{x_{k}}(T(z)-a_{i})dz\right]}| U_{R,i}(x_{k}) \rangle
     \\ 
     + \sum_{i=1}^{n} A_{i} \int_{x_{k-1}}^{x_{k}} dx \int_{x}^{x_{k}} dy  \langle U_{L,i}(x)| \nabla_{Q_{k}} Q(x)  P\exp{\left[\int_{x}^{y}(T(u)-a_{i}) du\right]}  R(y) \otimes \overline{R(y)} P\exp{\left[\int_{y}^{x_{k}}T(z)
    dz\right]}|\sigma_{R}(x_{k})\rangle 
    \\ 
    + \int_{x_{k-1}}^{x_{k}} dx \langle  \sigma_{L}(x)|\nabla_{Q_{k}} H(x)  P\exp{\left[\int_{x}^{x_{k}} T(z)dz\right]}|\sigma_{R}(x_{k})\rangle  + \int_{x_{k-1}}^{x_{k}} dx \langle \sigma_{L}(x)| \nabla_{Q_{k}} Q(x) P\exp{\left(\int_{x}^{x_{k}} T(u) du\right)} |H_{R}(x_{k})\rangle
    \\ 
    + \sum_{i=1}^{n} A_{i} \int_{x_{k-1}}^{x_{k}} dx \int_{x}^{x_{k}} dy  \langle \sigma_{L}(x)| \nabla_{Q_{k}} Q(x) P\exp{\left[\int_{x}^{y}T(u)du\right]}  R(y) \otimes \overline{R(y)} P\exp{\left[\int_{y}^{x_{k}}(T(z)-a_{i})
    dz\right]} |U_{R,i}(x_{k})\rangle 
    \\ 
    + \int_{x_{k-1}}^{x_{k}} dx \int_{x}^{x_{k}} dy  \langle \sigma_{L}(x)| \nabla_{Q_{k}} Q(x) P\exp{\left[\int_{x}^{y}T(u)du\right]}  H(y) P\exp{\left[\int_{y}^{x_{k}}T(z)
    dz\right]}|\sigma_{R}(x_{k})\rangle + K,
    \end{multline}
    where
    \begin{eqnarray*}
    K&=&
    \sum_{i=1}^{n} A_{i} \int_{x_{k-1}}^{x_{k}} dx \int_{x}^{x_{k}} dy \int_{y}^{x_{k}} dt \langle \sigma_{L}(x)| \nabla_{Q_{k}} Q(x) \times 
    \\ 
    &&\qquad\times P\exp{\left[\int_{x}^{y} T(v)
    dv\right]}  R(y) \otimes \overline{R(y)} P\exp{\left[\int_{y}^{t}(T(u)-a_{i}) du\right]}   R(t) \otimes \overline{R(t)} P\exp{\left[\int_{t}^{x_{k}}T(z)
    dz\right]} |\sigma_{R}(x_{k}) \rangle.
 \end{eqnarray*}
The obtained expression can be further simplified. For example, the term $$\int_{x_{k-1}}^{x_{k}} dx \langle  H_{L}(x)|\nabla_{Q_{k}} Q(x)  P\exp{\left[\int_{x}^{x_{k}} T(z)dz\right]}|\sigma_{R}(x_{k})\rangle = \int_{x_{k-1}}^{x_{k}} dx \langle  H_{L}(x)|\nabla_{Q_{k}} Q(x)  |\sigma_{R}(x)\rangle.$$
The second and third terms in Eq.~\eqref{DerEQ} can be expressed together as $\sum_{i=1}^{n} A_{i} \int_{x_{k-1}}^{x_{k}} dx \langle  U_{L,i}(x)|\nabla_{Q_{k}} Q(x)  |U_{R,i}(x)\rangle$, while the fourth term can be cast into the form $\int_{x_{k-1}}^{x_{k}} dx \langle  \sigma_{L}(x)|\nabla_{Q_{k}} H(x)  |\sigma_{R}(x)\rangle$. The last terms can be summed together into the integral of the form: $\int_{x_{k-1}}^{x_{k}} dx \langle  \sigma_{L}(x)|\nabla_{Q_{k}} Q(x)  |H_{R}(x)\rangle$. This can be explicitly verified by using the definitions of $H_{R}$ and $U_{R}$. 

By adding analogous terms with the derivatives of $\sigma_{R}(x_{k})$, $U_{R}(x_{k})$, and $H_{R}(x_{k})$, we obtain the following expression for the full derivative of the energy:
\begin{multline}\label{DEQ}
    \nabla_{Q_{k}} E/w = \int_{x_{k-1}}^{x_{k+1}} dx 
    {\langle  H_{L}(x)|\nabla_{Q_{k}} Q(x)  |\sigma_{R}(x)\rangle}
    +  \int_{x_{k-1}}^{x_{k+1}} dx 
    {\langle  \sigma_{L}(x)|\nabla_{Q_{k}} Q(x)  |H_{R}(x)\rangle}
    \\
    + \sum_{i=1}^{n} A_{i} \int_{x_{k-1}}^{x_{k+1}} dx 
    {\langle  U_{L,i}(x)|\nabla_{Q_{k}} Q(x)  |U_{R,i}(x)\rangle}
    +  \int_{x_{k-1}}^{x_{k+1}} dx
    {\langle  \sigma_{L}(x)|\left[
    \nabla_{Q_{k}} H(x)  - E \nabla_{Q_{k}} Q(x) 
    \right]
    |\sigma_{R}(x)\rangle}.
\end{multline}
The last term with the energy $E$ in Eq.~\eqref{DEQ} originates from the differentiation of the denominator in Eq.~\eqref{EnergyForGradient}. 

The full derivative of the energy by $R_{k}$ can be deduced in the same way, thus
\begin{multline}\label{DER}
    \nabla_{R_{k}} E/w = \int_{x_{k-1}}^{x_{k+1}} dx 
    {\langle  H_{L}(x)|\nabla_{R_{k}} R(x) \otimes \overline{R(x)} |\sigma_{R}(x)\rangle}
    + \int_{x_{k-1}}^{x_{k+1}} dx 
    {\langle  \sigma_{L}(x)| \nabla_{R_{k}} R(x) \otimes \overline{R(x)} |H_{R}(x)\rangle}
    \\
    + \sum_{i=1}^{n} A_{i} \int_{x_{k-1}}^{x_{k+1}} dx 
    \left\{
    {\langle  U_{L,i}(x)| \nabla_{R_{k}} R(x) \otimes \overline{R(x)} 
    \left[
        |U_{R,i}(x)\rangle + |\sigma_{R}(x_{k})\rangle
    \right]
    }
    + {\langle  \sigma_{L}(x)| \nabla_{R_{k}} R(x) \otimes \overline{R(x)} |U_{R,i}(x)\rangle}
    \right\}
    \\ 
    + \int_{x_{k-1}}^{x_{k+1}} dx 
    {\langle  \sigma_{L}(x)|
    \left[\nabla_{R_{k}} H(x)  - E\nabla_{R_{k}} R(x) \otimes \overline{R(x)}\right]|\sigma_{R}(x)\rangle}.
\end{multline}
\begin{figure}[t]
  \includegraphics[width=\linewidth]{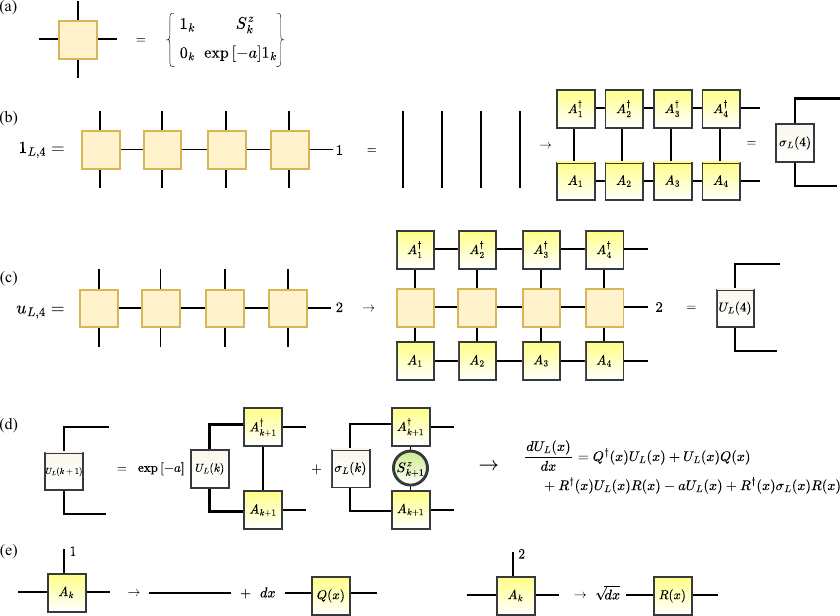}
  \caption{\label{fig:MPO}%
    Illustration of different stages of correspondence between the lattice MPO and the developed cMPS approach:
    (a) Definition of the MPO matrix, which encodes  the recursion relation~\eqref{recursionU} for $1_{L,k}$ and $u_{L,k}$;
    (b) MPO with the last index $1$ results in the identity operator $1_{L,k}$ (this operator can be sandwiched with the MPS wave function, which results in left density matrix $\sigma_{L}(k)$); 
    (c) MPO with the last index $2$ produces the operator $u_{L,k}$ (after sandwiching with the MPS wave function one obtains the matrix $U_{L}(k)$); 
    (d) recursion relation between $u_{L,k}$, $u_{L,k-1}$, and $1_{L,k-1}$ transforms into the recursion relation between the matrices $U_{L}(k+1)$, $U_{L}(k)$, and $\sigma_{L}(k)$; 
    after the continuous limit specified in (e), the recursion relation (d) becomes the Lindblad equation~\eqref{ULindbladCorrected}.
    }
\end{figure}
\end{widetext}

To evaluate the energy gradients with Eqs.~\eqref{DEQ} and \eqref{DER}, one needs to precompute the matrices $\sigma$, $H$, $U$ and then to calculate integrals explicitly with the help of the beta-functions, as described in Ref.~\cite{PhysRevLett.128.020501}.

The computational complexity of the energy and gradients calculation was partially discussed in Ref.~\cite{PhysRevLett.128.020501}. It scales as $D^{3} N_{mesh}$, similar to the DMRG case. The $D^{3}$ scaling originates from the matrix-matrix multiplications, while $N_{mesh}$ results from the number of coordinate intervals. The introduction of additional $n$ exponents in the long-range potential leads to the linear enhancement of the computational cost to $D^{3} N_{mesh}(2+n)$, which is also the same as in the usual DMRG computations with MPO of the bond dimension $(2+n)$. This increase can be partially reduced by parallelization of the calculations with different exponents, since these calculations are independent.

In conclusion, we can add several comments on the calculation of gradients. First, if matrices $H$, $U$, and $\sigma$ are provided, the gradients of the energy for different $k$ can be computed simultaneously in parallel.
Second, for a given $k$, the gradients depend only on the functional values determined in a small spatial range near $x_{k}$.
If one updates $R_{k}$, then the matrices $H$ and $\sigma$ will not change outside of this range (the same holds for the DMRG algorithm). 
In principle, one can use this fact to optimise the cMPS algorithm not globally, but locally in sweeps, as in the usual lattice DMRG (though here the local problem remains very challenging). 
This analogy with DMRG method is discussed in Appendix~\ref{AppB} in more detail. 
We leave the investigation of the cMPS optimization with local sweeps for a future research.

\section{Comparison with MPO methods on the lattice}\label{AppB}

Long-range interacting systems on the lattice can be simulated with DMRG using MPO, which efficiently encodes these interactions. 
We should note that if MPO has a small bond dimension $\chi$, then only two types of interactions can be efficiently encoded into MPO: (i) interactions, which exponentially decrease with distance~\cite{PhysRevB.78.035116}, and (ii) interactions in the cavity~\cite{Chanda2021SelforganizedTI, https://doi.org/10.48550/arxiv.2201.05466}. 
In Ref.~\cite{PhysRevB.78.035116}, it was proposed to approximate general interactions with a sum of exponents to encode them into MPO of a small bond dimension. This encoding was later implemented in various methods and applications. 

For a detailed comparison with the cMPS algorithm, let us illustrate the lattice MPO construction for the transverse Ising model with the exponentially decaying interaction amplitude. 
The model Hamiltonian is defined as follows:
\begin{equation}\label{ExpIsing}
    H_{\rm Ising} = \sum_{1\leq i<j}^L \exp{[-a(j-i-1)]}S_{i}^{z} S_{j}^{z} +g\sum_{i=1}^{L} S_{i}^{x}.
\end{equation}

The general MPO construction scheme is based on the decomposition of the Hamiltonian on three parts with the fixed bond ($k,k+1$) between the lattice sites $k$ and $k+1$,
\begin{equation}\label{eq:decomp}
    H = H_{L, k} \otimes 1_{R} + 1_{L} \otimes H_{R, k} + \sum_{m=1}^{N} u_{m,L,k} \otimes u_{m,R,k}.
\end{equation}
In this decomposition $H_{L,k}$ contains all operators acting on the sites to the left from the bond ($k,k+1$). 
For the Ising model~\eqref{ExpIsing}, it is expressed as
\begin{equation}\label{eq:HLk}
    H_{L,k} = g\sum_{i=1}^{k} S_{i}^{x} + \sum_{1\leq i < j}^k \exp{[-a(j-i-1)]}S_{i}^{z} S_{j}^{z}.
\end{equation}
Analogously, $H_{R,k}$ is acting on the sites to the right from the bond ($k,k+1$),
\begin{equation}
    H_{R,k} = g\sum_{i=k+1}^{L} S_{i}^{x} 
    + \sum_{k+1 \leq i < j}^L \exp{[-a(j-i-1)]}S_{i}^{z} S_{j}^{z}.
\end{equation}
These two operators have a similar role in DMRG to the density matrices $H_{L}(x)$ and $H_{R}(x)$ from Sec.~\ref{sec:method}. 
$u_{L,m,k}$ and $u_{R,m,k}$ are the operators acting separately to the left and to the right sides from the fixed bond, respectively, but their product is acting on both sides from the bond ($k,k+1$). 

For the transverse Ising model~\eqref{ExpIsing}, the parameter~$N$, which controls the bond dimension of the MPO in the sum~\eqref{eq:decomp}, is equal to one, therefore, 
\begin{eqnarray}\label{eq:uLk}
    &&u_{L,k} = \sum_{i=1}^{k} \exp{[-a (k-i)]} S_{i}^{z},
    \\
    &&u_{R,k} = \sum_{j=k+1}^{L} \exp{[-a (j-k-1)]} S_{j}^{z}.
    \nonumber
\end{eqnarray}
These operators are the lattice analogs of the matrices $ U_{L}(x)$ and $U_{R}(x)$, see Eqs.~\eqref{UL} and \eqref{UR}, respectively. 

The next step in the construction of MPO for the lattice Hamiltonian are the recursion relations, which allow us to express $H_{L,k}$ and $u_{L,k}$ on the bond ($k,k+1$) in terms of analogous operators on the bond ($k-1,k$). Let us start with derivation of the recursion relation for $u_{L,k}$.
Obviously, according to Eq.~\eqref{eq:uLk} we can express
\begin{eqnarray}\label{recursionU}
    u_{L,k} 
    &=&
    \exp{[-a]} u_{L,k-1} + S_{k}^{z}. 
\end{eqnarray}
This recursion relation can be viewed as the lattice version of Eq.~\eqref{ULindbladCorrected}, where the multiplication by $\exp{[-\mu]}$ is a lattice version of the term $-a U_{L,z}$ in the right-hand side of the differential equation for $U_{L}(z)$, while $S_{k}^{z}$ is analogous to $R^{\dagger}(z) \sigma_{L}(z) R(z)$.  
If we additionally introduce the identity operator $1_{L,k}$ acting on the first $k$ sites, we can rewrite the recursion relation~\eqref{recursionU} in the matrix-product form,
\begin{equation}
    (1_{L,k}, u_{L,k}) = (1_{L,k-1}, u_{L,k-1}) 
    \begin{pmatrix}
    1_{k} & S_{k}^{z} \\
    0_{k} & \exp{[-a]}1_{k}
    \end{pmatrix}.
\end{equation}

As a result of the repeatable application of this matrix-product recursion, we can rewrite the operators $1_{L,k}$ and $u_{L,k}$ in the form of MPO, as shown in Figs.~\subfigrefs{fig:MPO}{a}{c}.
Regarding the part $H_{L,k}$, which is given by Eq.~\eqref{eq:HLk}, we can similarly derive that
\begin{equation}
    H_{L,k} 
    = H_{L,k-1} + u_{L,k-1}S_{k}^{z} + g S_{k}^{x}.
\end{equation}
This recursion relation is a discrete analog of the Lindblad equation~\eqref{EnergyLindblad}. 
Here $gS_{k}^{x}$ is a discrete version of the local operator $H(x)$, while $u_{L,k-1}S_{k}^{z}$ is similar to the term $R^{\dagger}(z) U_{L}(z) R(z)$. The recursion for $H_{L,k}$ can be also rewritten in the matrix-product form. 

Note that certain discrepancies between the continuous and discrete systems still remain. 
First, the obtained equations in discrete systems are the recursion relations between operators, while the obtained Lindblad equations describe the density matrices. 
Second, the equations for discrete systems do not have any correspondence for the terms of the type $Q^{\dagger}(z) H_{L}(z) + H_{L}(z) Q(z) + R^{\dagger}(z) H_{L}(z) R(z)$ in the right-hand side of Eq.~\eqref{EnergyLindblad}. 
The latter discrepancies can be lifted by sandwiching the MPO operator between the MPS wave functions, as it is shown in Figs.~\subfigref{fig:MPO}{b} and \subfigref{fig:MPO}{c}.  
As a result of this procedure, we obtain the matrices $\sigma_{L}(k)$ and $U_{L}(k)$, as well as $H_{L}(k)$ (not shown in Fig.~\ref{fig:MPO}). 
The recursion relation between $u_{L,k}$ and $1_{L,k}$ translates into the recursion between the matrices $U_{L,k}$ and $\sigma_{L}(k)$, which is shown in Fig.~\subfigref{fig:MPO}{d}. 
In the last step, one can take the continuous limit of the MPS wave function [shown in Fig.~\subfigref{fig:MPO}{e}] to obtain the inhomogeneous Lindblad equation for $U_{L}(x)$, which is a continuous limit of $U_{L}(k)$. 
The terms in the Lindblad equation of the form $Q^{\dagger}(x) U_{L}(x)$ are obtained from the continuous limit of the MPS wave function. 

In the numerical procedure, the matrices of the form $\sigma_{L}(k)$, $U_{L}(k)$, and $H_{L}(k)$ are calculated and kept in the computer memory (with updates during the sweep) in the course of the DMRG algorithm. In the continuous case, we can also propagate these matrices with the Lindblad equation back and forth during the sweep through the coordinate interval, with the sequential update of $R_{k}$ and $Q_{k}$, using only the local gradients (which can be computed using only $\sigma_{L}(x)$, $U_{L}(x)$, and $H_{L}(x)$ in the proximity of $x=x_{k}$, as it is shown in Appendix~\ref{AppA}). 
We do not perform this sequential update within this study, but this is an interesting possibility, since the DMRG optimization by sweeps is very effective in the lattice case.

\bibliography{cMPS}

\begin{thebibliography}{49}%
\makeatletter
\providecommand \@ifxundefined [1]{%
 \@ifx{#1\undefined}
}%
\providecommand \@ifnum [1]{%
 \ifnum #1\expandafter \@firstoftwo
 \else \expandafter \@secondoftwo
 \fi
}%
\providecommand \@ifx [1]{%
 \ifx #1\expandafter \@firstoftwo
 \else \expandafter \@secondoftwo
 \fi
}%
\providecommand \natexlab [1]{#1}%
\providecommand \enquote  [1]{``#1''}%
\providecommand \bibnamefont  [1]{#1}%
\providecommand \bibfnamefont [1]{#1}%
\providecommand \citenamefont [1]{#1}%
\providecommand \href@noop [0]{\@secondoftwo}%
\providecommand \href [0]{\begingroup \@sanitize@url \@href}%
\providecommand \@href[1]{\@@startlink{#1}\@@href}%
\providecommand \@@href[1]{\endgroup#1\@@endlink}%
\providecommand \@sanitize@url [0]{\catcode `\\12\catcode `\$12\catcode
  `\&12\catcode `\#12\catcode `\^12\catcode `\_12\catcode `\%12\relax}%
\providecommand \@@startlink[1]{}%
\providecommand \@@endlink[0]{}%
\providecommand \url  [0]{\begingroup\@sanitize@url \@url }%
\providecommand \@url [1]{\endgroup\@href {#1}{\urlprefix }}%
\providecommand \urlprefix  [0]{URL }%
\providecommand \Eprint [0]{\href }%
\providecommand \doibase [0]{http://dx.doi.org/}%
\providecommand \selectlanguage [0]{\@gobble}%
\providecommand \bibinfo  [0]{\@secondoftwo}%
\providecommand \bibfield  [0]{\@secondoftwo}%
\providecommand \translation [1]{[#1]}%
\providecommand \BibitemOpen [0]{}%
\providecommand \bibitemStop [0]{}%
\providecommand \bibitemNoStop [0]{.\EOS\space}%
\providecommand \EOS [0]{\spacefactor3000\relax}%
\providecommand \BibitemShut  [1]{\csname bibitem#1\endcsname}%
\let\auto@bib@innerbib\@empty
\bibitem [{\citenamefont {Schollwöck}(2011)}]{Schollwoeck2011TheDR}%
  \BibitemOpen
  \bibfield  {author} {\bibinfo {author} {\bibfnamefont {U.}~\bibnamefont
  {Schollwöck}},\ }\href {\doibase 10.1016/j.aop.2010.09.012} {\bibfield
  {journal} {\bibinfo  {journal} {Ann. Phys.}\ }\textbf {\bibinfo {volume}
  {326}},\ \bibinfo {pages} {96} (\bibinfo {year} {2011})}\BibitemShut
  {NoStop}%
\bibitem [{\citenamefont {Eisert}\ \emph {et~al.}(2010)\citenamefont {Eisert},
  \citenamefont {Cramer},\ and\ \citenamefont {Plenio}}]{Eisert2010AreaLF}%
  \BibitemOpen
  \bibfield  {author} {\bibinfo {author} {\bibfnamefont {J.}~\bibnamefont
  {Eisert}}, \bibinfo {author} {\bibfnamefont {M.}~\bibnamefont {Cramer}}, \
  and\ \bibinfo {author} {\bibfnamefont {M.~B.}\ \bibnamefont {Plenio}},\
  }\href {\doibase 10.1103/revmodphys.82.277} {\bibfield  {journal} {\bibinfo
  {journal} {Rev. Mod. Phys.}\ }\textbf {\bibinfo {volume} {82}},\ \bibinfo
  {pages} {277} (\bibinfo {year} {2010})}\BibitemShut {NoStop}%
\bibitem [{\citenamefont {Vidal}(2008)}]{MERA}%
  \BibitemOpen
  \bibfield  {author} {\bibinfo {author} {\bibfnamefont {G.}~\bibnamefont
  {Vidal}},\ }\href {\doibase 10.1103/PhysRevLett.101.110501} {\bibfield
  {journal} {\bibinfo  {journal} {Phys. Rev. Lett.}\ }\textbf {\bibinfo
  {volume} {101}},\ \bibinfo {pages} {110501} (\bibinfo {year}
  {2008})}\BibitemShut {NoStop}%
\bibitem [{\citenamefont {Verstraete}\ and\ \citenamefont
  {Cirac}(2004)}]{Verstraete2004RenormalizationAF}%
  \BibitemOpen
  \bibfield  {author} {\bibinfo {author} {\bibfnamefont {F.}~\bibnamefont
  {Verstraete}}\ and\ \bibinfo {author} {\bibfnamefont {J.~I.}\ \bibnamefont
  {Cirac}},\ }\href@noop {} {\enquote {\bibinfo {title} {Renormalization
  algorithms for quantum-many body systems in two and higher dimensions},}\ }
  (\bibinfo {year} {2004}),\ \Eprint {http://arxiv.org/abs/cond-mat/0407066}
  {arXiv:cond-mat/0407066} \BibitemShut {NoStop}%
\bibitem [{\citenamefont {Dolfi}\ \emph {et~al.}(2012)\citenamefont {Dolfi},
  \citenamefont {Bauer}, \citenamefont {Troyer},\ and\ \citenamefont
  {Ristivojevic}}]{MultigridMethods}%
  \BibitemOpen
  \bibfield  {author} {\bibinfo {author} {\bibfnamefont {M.}~\bibnamefont
  {Dolfi}}, \bibinfo {author} {\bibfnamefont {B.}~\bibnamefont {Bauer}},
  \bibinfo {author} {\bibfnamefont {M.}~\bibnamefont {Troyer}}, \ and\ \bibinfo
  {author} {\bibfnamefont {Z.}~\bibnamefont {Ristivojevic}},\ }\href {\doibase
  10.1103/PhysRevLett.109.020604} {\bibfield  {journal} {\bibinfo  {journal}
  {Phys. Rev. Lett.}\ }\textbf {\bibinfo {volume} {109}},\ \bibinfo {pages}
  {020604} (\bibinfo {year} {2012})}\BibitemShut {NoStop}%
\bibitem [{\citenamefont {Dutta}\ \emph {et~al.}(2022)\citenamefont {Dutta},
  \citenamefont {Buyskikh}, \citenamefont {Daley},\ and\ \citenamefont
  {Mueller}}]{Dutta2021DensityMatrixRG}%
  \BibitemOpen
  \bibfield  {author} {\bibinfo {author} {\bibfnamefont {S.}~\bibnamefont
  {Dutta}}, \bibinfo {author} {\bibfnamefont {A.}~\bibnamefont {Buyskikh}},
  \bibinfo {author} {\bibfnamefont {A.~J.}\ \bibnamefont {Daley}}, \ and\
  \bibinfo {author} {\bibfnamefont {E.~J.}\ \bibnamefont {Mueller}},\ }\href
  {\doibase 10.1103/PhysRevLett.128.230401} {\bibfield  {journal} {\bibinfo
  {journal} {Phys. Rev. Lett.}\ }\textbf {\bibinfo {volume} {128}},\ \bibinfo
  {pages} {230401} (\bibinfo {year} {2022})}\BibitemShut {NoStop}%
\bibitem [{\citenamefont {Ganahl}\ \emph {et~al.}(2017)\citenamefont {Ganahl},
  \citenamefont {Rinc\'on},\ and\ \citenamefont {Vidal}}]{cMPSfinegrain}%
  \BibitemOpen
  \bibfield  {author} {\bibinfo {author} {\bibfnamefont {M.}~\bibnamefont
  {Ganahl}}, \bibinfo {author} {\bibfnamefont {J.}~\bibnamefont {Rinc\'on}}, \
  and\ \bibinfo {author} {\bibfnamefont {G.}~\bibnamefont {Vidal}},\ }\href
  {\doibase 10.1103/PhysRevLett.118.220402} {\bibfield  {journal} {\bibinfo
  {journal} {Phys. Rev. Lett.}\ }\textbf {\bibinfo {volume} {118}},\ \bibinfo
  {pages} {220402} (\bibinfo {year} {2017})}\BibitemShut {NoStop}%
\bibitem [{\citenamefont {Ba{\~{n}}uls}\ \emph {et~al.}(2013)\citenamefont
  {Ba{\~{n}}uls}, \citenamefont {Cichy}, \citenamefont {Cirac},\ and\
  \citenamefont {Jansen}}]{Banuls_2013}%
  \BibitemOpen
  \bibfield  {author} {\bibinfo {author} {\bibfnamefont {M.}~\bibnamefont
  {Ba{\~{n}}uls}}, \bibinfo {author} {\bibfnamefont {K.}~\bibnamefont {Cichy}},
  \bibinfo {author} {\bibfnamefont {J.}~\bibnamefont {Cirac}}, \ and\ \bibinfo
  {author} {\bibfnamefont {K.}~\bibnamefont {Jansen}},\ }\href
  {https://doi.org/10.1007/jhep11(2013)158} {\bibfield  {journal} {\bibinfo
  {journal} {J. High Energ. Phys.}\ }\textbf {\bibinfo {volume} {2013}},\
  \bibinfo {pages} {158} (\bibinfo {year} {2013})}\BibitemShut {NoStop}%
\bibitem [{\citenamefont {Verstraete}\ and\ \citenamefont
  {Cirac}(2010)}]{cMPSIntro}%
  \BibitemOpen
  \bibfield  {author} {\bibinfo {author} {\bibfnamefont {F.}~\bibnamefont
  {Verstraete}}\ and\ \bibinfo {author} {\bibfnamefont {J.~I.}\ \bibnamefont
  {Cirac}},\ }\href {\doibase 10.1103/PhysRevLett.104.190405} {\bibfield
  {journal} {\bibinfo  {journal} {Phys. Rev. Lett.}\ }\textbf {\bibinfo
  {volume} {104}},\ \bibinfo {pages} {190405} (\bibinfo {year}
  {2010})}\BibitemShut {NoStop}%
\bibitem [{\citenamefont {Haegeman}\ \emph
  {et~al.}(2013{\natexlab{a}})\citenamefont {Haegeman}, \citenamefont {Cirac},
  \citenamefont {Osborne},\ and\ \citenamefont
  {Verstraete}}]{Haegeman2013CalculusOC}%
  \BibitemOpen
  \bibfield  {author} {\bibinfo {author} {\bibfnamefont {J.}~\bibnamefont
  {Haegeman}}, \bibinfo {author} {\bibfnamefont {J.~I.}\ \bibnamefont {Cirac}},
  \bibinfo {author} {\bibfnamefont {T.~J.}\ \bibnamefont {Osborne}}, \ and\
  \bibinfo {author} {\bibfnamefont {F.}~\bibnamefont {Verstraete}},\ }\href
  {\doibase 10.1103/PhysRevB.88.085118} {\bibfield  {journal} {\bibinfo
  {journal} {Phys. Rev. B}\ }\textbf {\bibinfo {volume} {88}},\ \bibinfo
  {pages} {085118} (\bibinfo {year} {2013}{\natexlab{a}})}\BibitemShut
  {NoStop}%
\bibitem [{\citenamefont {Ganahl}\ and\ \citenamefont
  {Vidal}(2018)}]{PhysRevB.98.195105}%
  \BibitemOpen
  \bibfield  {author} {\bibinfo {author} {\bibfnamefont {M.}~\bibnamefont
  {Ganahl}}\ and\ \bibinfo {author} {\bibfnamefont {G.}~\bibnamefont {Vidal}},\
  }\href {\doibase 10.1103/PhysRevB.98.195105} {\bibfield  {journal} {\bibinfo
  {journal} {Phys. Rev. B}\ }\textbf {\bibinfo {volume} {98}},\ \bibinfo
  {pages} {195105} (\bibinfo {year} {2018})}\BibitemShut {NoStop}%
\bibitem [{\citenamefont {Haegeman}\ \emph {et~al.}(2010)\citenamefont
  {Haegeman}, \citenamefont {Cirac}, \citenamefont {Osborne}, \citenamefont
  {Verschelde},\ and\ \citenamefont {Verstraete}}]{TransInvcMPS1}%
  \BibitemOpen
  \bibfield  {author} {\bibinfo {author} {\bibfnamefont {J.}~\bibnamefont
  {Haegeman}}, \bibinfo {author} {\bibfnamefont {J.~I.}\ \bibnamefont {Cirac}},
  \bibinfo {author} {\bibfnamefont {T.~J.}\ \bibnamefont {Osborne}}, \bibinfo
  {author} {\bibfnamefont {H.}~\bibnamefont {Verschelde}}, \ and\ \bibinfo
  {author} {\bibfnamefont {F.}~\bibnamefont {Verstraete}},\ }\href {\doibase
  10.1103/PhysRevLett.105.251601} {\bibfield  {journal} {\bibinfo  {journal}
  {Phys. Rev. Lett.}\ }\textbf {\bibinfo {volume} {105}},\ \bibinfo {pages}
  {251601} (\bibinfo {year} {2010})}\BibitemShut {NoStop}%
\bibitem [{\citenamefont {Vanderstraeten}\ \emph {et~al.}(2019)\citenamefont
  {Vanderstraeten}, \citenamefont {Haegeman},\ and\ \citenamefont
  {Verstraete}}]{Vanderstraeten2019TangentspaceMF}%
  \BibitemOpen
  \bibfield  {author} {\bibinfo {author} {\bibfnamefont {L.}~\bibnamefont
  {Vanderstraeten}}, \bibinfo {author} {\bibfnamefont {J.}~\bibnamefont
  {Haegeman}}, \ and\ \bibinfo {author} {\bibfnamefont {F.}~\bibnamefont
  {Verstraete}},\ }\href {\doibase 10.21468/SciPostPhysLectNotes.7} {\bibfield
  {journal} {\bibinfo  {journal} {SciPost Phys. Lect. Notes}\ }\textbf
  {\bibinfo {volume} {7}},\ \bibinfo {pages} {1} (\bibinfo {year}
  {2019})}\BibitemShut {NoStop}%
\bibitem [{\citenamefont {Haegeman}\ \emph {et~al.}(2017)\citenamefont
  {Haegeman}, \citenamefont {Draxler}, \citenamefont {Stojevic}, \citenamefont
  {Cirac}, \citenamefont {Osborne},\ and\ \citenamefont
  {Verstraete}}]{SciPostPhys.3.1.006}%
  \BibitemOpen
  \bibfield  {author} {\bibinfo {author} {\bibfnamefont {J.}~\bibnamefont
  {Haegeman}}, \bibinfo {author} {\bibfnamefont {D.}~\bibnamefont {Draxler}},
  \bibinfo {author} {\bibfnamefont {V.}~\bibnamefont {Stojevic}}, \bibinfo
  {author} {\bibfnamefont {J.~I.}\ \bibnamefont {Cirac}}, \bibinfo {author}
  {\bibfnamefont {T.~J.}\ \bibnamefont {Osborne}}, \ and\ \bibinfo {author}
  {\bibfnamefont {F.}~\bibnamefont {Verstraete}},\ }\href {\doibase
  10.21468/SciPostPhys.3.1.006} {\bibfield  {journal} {\bibinfo  {journal}
  {SciPost Phys.}\ }\textbf {\bibinfo {volume} {3}},\ \bibinfo {pages} {006}
  (\bibinfo {year} {2017})}\BibitemShut {NoStop}%
\bibitem [{\citenamefont {Rinc\'on}\ \emph {et~al.}(2015)\citenamefont
  {Rinc\'on}, \citenamefont {Ganahl},\ and\ \citenamefont
  {Vidal}}]{PhysRevB.92.115107}%
  \BibitemOpen
  \bibfield  {author} {\bibinfo {author} {\bibfnamefont {J.}~\bibnamefont
  {Rinc\'on}}, \bibinfo {author} {\bibfnamefont {M.}~\bibnamefont {Ganahl}}, \
  and\ \bibinfo {author} {\bibfnamefont {G.}~\bibnamefont {Vidal}},\ }\href
  {\doibase 10.1103/PhysRevB.92.115107} {\bibfield  {journal} {\bibinfo
  {journal} {Phys. Rev. B}\ }\textbf {\bibinfo {volume} {92}},\ \bibinfo
  {pages} {115107} (\bibinfo {year} {2015})}\BibitemShut {NoStop}%
\bibitem [{\citenamefont {Tilloy}(2021)}]{PhysRevD.104.096007}%
  \BibitemOpen
  \bibfield  {author} {\bibinfo {author} {\bibfnamefont {A.}~\bibnamefont
  {Tilloy}},\ }\href {\doibase 10.1103/PhysRevD.104.096007} {\bibfield
  {journal} {\bibinfo  {journal} {Phys. Rev. D}\ }\textbf {\bibinfo {volume}
  {104}},\ \bibinfo {pages} {096007} (\bibinfo {year} {2021})}\BibitemShut
  {NoStop}%
\bibitem [{\citenamefont {Draxler}\ \emph {et~al.}(2017)\citenamefont
  {Draxler}, \citenamefont {Haegeman}, \citenamefont {Verstraete},\ and\
  \citenamefont {Rizzi}}]{TimeDynamics}%
  \BibitemOpen
  \bibfield  {author} {\bibinfo {author} {\bibfnamefont {D.}~\bibnamefont
  {Draxler}}, \bibinfo {author} {\bibfnamefont {J.}~\bibnamefont {Haegeman}},
  \bibinfo {author} {\bibfnamefont {F.}~\bibnamefont {Verstraete}}, \ and\
  \bibinfo {author} {\bibfnamefont {M.}~\bibnamefont {Rizzi}},\ }\href
  {\doibase 10.1103/PhysRevB.95.045145} {\bibfield  {journal} {\bibinfo
  {journal} {Phys. Rev. B}\ }\textbf {\bibinfo {volume} {95}},\ \bibinfo
  {pages} {045145} (\bibinfo {year} {2017})}\BibitemShut {NoStop}%
\bibitem [{\citenamefont {Jennings}\ \emph {et~al.}(2015)\citenamefont
  {Jennings}, \citenamefont {Brockt}, \citenamefont {Haegeman}, \citenamefont
  {Osborne},\ and\ \citenamefont {Verstraete}}]{Jennings_2015}%
  \BibitemOpen
  \bibfield  {author} {\bibinfo {author} {\bibfnamefont {D.}~\bibnamefont
  {Jennings}}, \bibinfo {author} {\bibfnamefont {C.}~\bibnamefont {Brockt}},
  \bibinfo {author} {\bibfnamefont {J.}~\bibnamefont {Haegeman}}, \bibinfo
  {author} {\bibfnamefont {T.~J.}\ \bibnamefont {Osborne}}, \ and\ \bibinfo
  {author} {\bibfnamefont {F.}~\bibnamefont {Verstraete}},\ }\href {\doibase
  10.1088/1367-2630/17/6/063039} {\bibfield  {journal} {\bibinfo  {journal}
  {New J. Phys.}\ }\textbf {\bibinfo {volume} {17}},\ \bibinfo {pages} {063039}
  (\bibinfo {year} {2015})}\BibitemShut {NoStop}%
\bibitem [{\citenamefont {Tilloy}\ and\ \citenamefont {Cirac}(2019)}]{cPEPS}%
  \BibitemOpen
  \bibfield  {author} {\bibinfo {author} {\bibfnamefont {A.}~\bibnamefont
  {Tilloy}}\ and\ \bibinfo {author} {\bibfnamefont {J.~I.}\ \bibnamefont
  {Cirac}},\ }\href {\doibase 10.1103/PhysRevX.9.021040} {\bibfield  {journal}
  {\bibinfo  {journal} {Phys. Rev. X}\ }\textbf {\bibinfo {volume} {9}},\
  \bibinfo {pages} {021040} (\bibinfo {year} {2019})}\BibitemShut {NoStop}%
\bibitem [{\citenamefont {Shachar}\ and\ \citenamefont {Zohar}(2022)}]{cPEPS2}%
  \BibitemOpen
  \bibfield  {author} {\bibinfo {author} {\bibfnamefont {T.}~\bibnamefont
  {Shachar}}\ and\ \bibinfo {author} {\bibfnamefont {E.}~\bibnamefont
  {Zohar}},\ }\href {\doibase 10.1103/PhysRevD.105.045016} {\bibfield
  {journal} {\bibinfo  {journal} {Phys. Rev. D}\ }\textbf {\bibinfo {volume}
  {105}},\ \bibinfo {pages} {045016} (\bibinfo {year} {2022})}\BibitemShut
  {NoStop}%
\bibitem [{\citenamefont {Haegeman}\ \emph
  {et~al.}(2013{\natexlab{b}})\citenamefont {Haegeman}, \citenamefont
  {Osborne}, \citenamefont {Verschelde},\ and\ \citenamefont
  {Verstraete}}]{cMERA}%
  \BibitemOpen
  \bibfield  {author} {\bibinfo {author} {\bibfnamefont {J.}~\bibnamefont
  {Haegeman}}, \bibinfo {author} {\bibfnamefont {T.~J.}\ \bibnamefont
  {Osborne}}, \bibinfo {author} {\bibfnamefont {H.}~\bibnamefont {Verschelde}},
  \ and\ \bibinfo {author} {\bibfnamefont {F.}~\bibnamefont {Verstraete}},\
  }\href {\doibase 10.1103/PhysRevLett.110.100402} {\bibfield  {journal}
  {\bibinfo  {journal} {Phys. Rev. Lett.}\ }\textbf {\bibinfo {volume} {110}},\
  \bibinfo {pages} {100402} (\bibinfo {year} {2013}{\natexlab{b}})}\BibitemShut
  {NoStop}%
\bibitem [{\citenamefont {Tang}\ \emph {et~al.}(2020)\citenamefont {Tang},
  \citenamefont {Tu},\ and\ \citenamefont {Wang}}]{cMPO}%
  \BibitemOpen
  \bibfield  {author} {\bibinfo {author} {\bibfnamefont {W.}~\bibnamefont
  {Tang}}, \bibinfo {author} {\bibfnamefont {H.-H.}\ \bibnamefont {Tu}}, \ and\
  \bibinfo {author} {\bibfnamefont {L.}~\bibnamefont {Wang}},\ }\href {\doibase
  10.1103/PhysRevLett.125.170604} {\bibfield  {journal} {\bibinfo  {journal}
  {Phys. Rev. Lett.}\ }\textbf {\bibinfo {volume} {125}},\ \bibinfo {pages}
  {170604} (\bibinfo {year} {2020})}\BibitemShut {NoStop}%
\bibitem [{\citenamefont {Osborne}\ \emph {et~al.}(2010)\citenamefont
  {Osborne}, \citenamefont {Eisert},\ and\ \citenamefont
  {Verstraete}}]{PhysRevLett.105.260401}%
  \BibitemOpen
  \bibfield  {author} {\bibinfo {author} {\bibfnamefont {T.~J.}\ \bibnamefont
  {Osborne}}, \bibinfo {author} {\bibfnamefont {J.}~\bibnamefont {Eisert}}, \
  and\ \bibinfo {author} {\bibfnamefont {F.}~\bibnamefont {Verstraete}},\
  }\href {\doibase 10.1103/PhysRevLett.105.260401} {\bibfield  {journal}
  {\bibinfo  {journal} {Phys. Rev. Lett.}\ }\textbf {\bibinfo {volume} {105}},\
  \bibinfo {pages} {260401} (\bibinfo {year} {2010})}\BibitemShut {NoStop}%
\bibitem [{\citenamefont {Kiukas}\ \emph {et~al.}(2015)\citenamefont {Kiukas},
  \citenamefont {Gu\ifmmode \mbox{\c{t}}\else \c{t}\fi{}\ifmmode~\u{a}\else
  \u{a}\fi{}}, \citenamefont {Lesanovsky},\ and\ \citenamefont
  {Garrahan}}]{OpenSystems}%
  \BibitemOpen
  \bibfield  {author} {\bibinfo {author} {\bibfnamefont {J.}~\bibnamefont
  {Kiukas}}, \bibinfo {author} {\bibfnamefont {M.}~\bibnamefont {Gu\ifmmode
  \mbox{\c{t}}\else \c{t}\fi{}\ifmmode~\u{a}\else \u{a}\fi{}}}, \bibinfo
  {author} {\bibfnamefont {I.}~\bibnamefont {Lesanovsky}}, \ and\ \bibinfo
  {author} {\bibfnamefont {J.~P.}\ \bibnamefont {Garrahan}},\ }\href {\doibase
  10.1103/PhysRevE.92.012132} {\bibfield  {journal} {\bibinfo  {journal} {Phys.
  Rev. E}\ }\textbf {\bibinfo {volume} {92}},\ \bibinfo {pages} {012132}
  (\bibinfo {year} {2015})}\BibitemShut {NoStop}%
\bibitem [{\citenamefont {Garrahan}(2016)}]{Garrahan_2016}%
  \BibitemOpen
  \bibfield  {author} {\bibinfo {author} {\bibfnamefont {J.~P.}\ \bibnamefont
  {Garrahan}},\ }\href {\doibase 10.1088/1742-5468/2016/07/073208} {\bibfield
  {journal} {\bibinfo  {journal} {J. Stat. Mech.}\ }\textbf {\bibinfo {volume}
  {2016}},\ \bibinfo {pages} {073208} (\bibinfo {year} {2016})}\BibitemShut
  {NoStop}%
\bibitem [{\citenamefont {Ganahl}(2017)}]{Ganahl2017ContinuousMP}%
  \BibitemOpen
  \bibfield  {author} {\bibinfo {author} {\bibfnamefont {M.}~\bibnamefont
  {Ganahl}},\ }\href@noop {} {\enquote {\bibinfo {title} {Continuous matrix
  product states for inhomogeneous quantum field theories: a basis-spline
  approach},}\ } (\bibinfo {year} {2017}),\ \Eprint
  {http://arxiv.org/abs/1712.01260} {arXiv:1712.01260} \BibitemShut {NoStop}%
\bibitem [{\citenamefont {Tuybens}\ \emph {et~al.}(2022)\citenamefont
  {Tuybens}, \citenamefont {De~Nardis}, \citenamefont {Haegeman},\ and\
  \citenamefont {Verstraete}}]{PhysRevLett.128.020501}%
  \BibitemOpen
  \bibfield  {author} {\bibinfo {author} {\bibfnamefont {B.}~\bibnamefont
  {Tuybens}}, \bibinfo {author} {\bibfnamefont {J.}~\bibnamefont {De~Nardis}},
  \bibinfo {author} {\bibfnamefont {J.}~\bibnamefont {Haegeman}}, \ and\
  \bibinfo {author} {\bibfnamefont {F.}~\bibnamefont {Verstraete}},\ }\href
  {\doibase 10.1103/PhysRevLett.128.020501} {\bibfield  {journal} {\bibinfo
  {journal} {Phys. Rev. Lett.}\ }\textbf {\bibinfo {volume} {128}},\ \bibinfo
  {pages} {020501} (\bibinfo {year} {2022})}\BibitemShut {NoStop}%
\bibitem [{\citenamefont {Chung}\ \emph {et~al.}(2015)\citenamefont {Chung},
  \citenamefont {Sun},\ and\ \citenamefont {Bolech}}]{cMPSFermions}%
  \BibitemOpen
  \bibfield  {author} {\bibinfo {author} {\bibfnamefont {S.~S.}\ \bibnamefont
  {Chung}}, \bibinfo {author} {\bibfnamefont {K.}~\bibnamefont {Sun}}, \ and\
  \bibinfo {author} {\bibfnamefont {C.~J.}\ \bibnamefont {Bolech}},\ }\href
  {\doibase 10.1103/PhysRevB.91.121108} {\bibfield  {journal} {\bibinfo
  {journal} {Phys. Rev. B}\ }\textbf {\bibinfo {volume} {91}},\ \bibinfo
  {pages} {121108} (\bibinfo {year} {2015})}\BibitemShut {NoStop}%
\bibitem [{\citenamefont {Quijandr\'{\i}a}\ \emph {et~al.}(2014)\citenamefont
  {Quijandr\'{\i}a}, \citenamefont {Garc\'{\i}a-Ripoll},\ and\ \citenamefont
  {Zueco}}]{Mixtures}%
  \BibitemOpen
  \bibfield  {author} {\bibinfo {author} {\bibfnamefont {F.}~\bibnamefont
  {Quijandr\'{\i}a}}, \bibinfo {author} {\bibfnamefont {J.~J.}\ \bibnamefont
  {Garc\'{\i}a-Ripoll}}, \ and\ \bibinfo {author} {\bibfnamefont
  {D.}~\bibnamefont {Zueco}},\ }\href {\doibase 10.1103/PhysRevB.90.235142}
  {\bibfield  {journal} {\bibinfo  {journal} {Phys. Rev. B}\ }\textbf {\bibinfo
  {volume} {90}},\ \bibinfo {pages} {235142} (\bibinfo {year}
  {2014})}\BibitemShut {NoStop}%
\bibitem [{\citenamefont {Chung}\ and\ \citenamefont
  {Bolech}(2017)}]{cMPSmixtures}%
  \BibitemOpen
  \bibfield  {author} {\bibinfo {author} {\bibfnamefont {S.~S.}\ \bibnamefont
  {Chung}}\ and\ \bibinfo {author} {\bibfnamefont {C.~J.}\ \bibnamefont
  {Bolech}},\ }\href {\doibase 10.1103/PhysRevA.96.023609} {\bibfield
  {journal} {\bibinfo  {journal} {Phys. Rev. A}\ }\textbf {\bibinfo {volume}
  {96}},\ \bibinfo {pages} {023609} (\bibinfo {year} {2017})}\BibitemShut
  {NoStop}%
\bibitem [{\citenamefont {Quijandr\'{\i}a}\ and\ \citenamefont
  {Zueco}(2015)}]{BBMixtures}%
  \BibitemOpen
  \bibfield  {author} {\bibinfo {author} {\bibfnamefont {F.}~\bibnamefont
  {Quijandr\'{\i}a}}\ and\ \bibinfo {author} {\bibfnamefont {D.}~\bibnamefont
  {Zueco}},\ }\href {\doibase 10.1103/PhysRevA.92.043629} {\bibfield  {journal}
  {\bibinfo  {journal} {Phys. Rev. A}\ }\textbf {\bibinfo {volume} {92}},\
  \bibinfo {pages} {043629} (\bibinfo {year} {2015})}\BibitemShut {NoStop}%
\bibitem [{\citenamefont {Peacock}\ \emph {et~al.}(2022)\citenamefont
  {Peacock}, \citenamefont {Ljepoja},\ and\ \citenamefont
  {Bolech}}]{BFMixtures}%
  \BibitemOpen
  \bibfield  {author} {\bibinfo {author} {\bibfnamefont {J.~C.}\ \bibnamefont
  {Peacock}}, \bibinfo {author} {\bibfnamefont {A.}~\bibnamefont {Ljepoja}}, \
  and\ \bibinfo {author} {\bibfnamefont {C.~J.}\ \bibnamefont {Bolech}},\
  }\href {\doibase 10.1103/PhysRevResearch.4.L022034} {\bibfield  {journal}
  {\bibinfo  {journal} {Phys. Rev. Research}\ }\textbf {\bibinfo {volume}
  {4}},\ \bibinfo {pages} {L022034} (\bibinfo {year} {2022})}\BibitemShut
  {NoStop}%
\bibitem [{\citenamefont {Mottl}\ \emph {et~al.}(2012)\citenamefont {Mottl},
  \citenamefont {Brennecke}, \citenamefont {Baumann}, \citenamefont {Landig},
  \citenamefont {Donner},\ and\ \citenamefont
  {Esslinger}}]{Mottl2014RotontypeMS}%
  \BibitemOpen
  \bibfield  {author} {\bibinfo {author} {\bibfnamefont {R.}~\bibnamefont
  {Mottl}}, \bibinfo {author} {\bibfnamefont {F.}~\bibnamefont {Brennecke}},
  \bibinfo {author} {\bibfnamefont {K.}~\bibnamefont {Baumann}}, \bibinfo
  {author} {\bibfnamefont {R.}~\bibnamefont {Landig}}, \bibinfo {author}
  {\bibfnamefont {T.}~\bibnamefont {Donner}}, \ and\ \bibinfo {author}
  {\bibfnamefont {T.}~\bibnamefont {Esslinger}},\ }\href {\doibase
  10.1126/science.1220314} {\bibfield  {journal} {\bibinfo  {journal}
  {Science}\ }\textbf {\bibinfo {volume} {336}},\ \bibinfo {pages} {1570}
  (\bibinfo {year} {2012})}\BibitemShut {NoStop}%
\bibitem [{\citenamefont {Gopalakrishnan}\ \emph {et~al.}(2009)\citenamefont
  {Gopalakrishnan}, \citenamefont {Lev},\ and\ \citenamefont
  {Goldbart}}]{Gopalakrishnan_2009}%
  \BibitemOpen
  \bibfield  {author} {\bibinfo {author} {\bibfnamefont {S.}~\bibnamefont
  {Gopalakrishnan}}, \bibinfo {author} {\bibfnamefont {B.~L.}\ \bibnamefont
  {Lev}}, \ and\ \bibinfo {author} {\bibfnamefont {P.~M.}\ \bibnamefont
  {Goldbart}},\ }\href {\doibase 10.1038/nphys1403} {\bibfield  {journal}
  {\bibinfo  {journal} {Nat. Phys.}\ }\textbf {\bibinfo {volume} {5}},\
  \bibinfo {pages} {845} (\bibinfo {year} {2009})}\BibitemShut {NoStop}%
\bibitem [{\citenamefont {Crosswhite}\ \emph {et~al.}(2008)\citenamefont
  {Crosswhite}, \citenamefont {Doherty},\ and\ \citenamefont
  {Vidal}}]{PhysRevB.78.035116}%
  \BibitemOpen
  \bibfield  {author} {\bibinfo {author} {\bibfnamefont {G.~M.}\ \bibnamefont
  {Crosswhite}}, \bibinfo {author} {\bibfnamefont {A.~C.}\ \bibnamefont
  {Doherty}}, \ and\ \bibinfo {author} {\bibfnamefont {G.}~\bibnamefont
  {Vidal}},\ }\href {\doibase 10.1103/PhysRevB.78.035116} {\bibfield  {journal}
  {\bibinfo  {journal} {Phys. Rev. B}\ }\textbf {\bibinfo {volume} {78}},\
  \bibinfo {pages} {035116} (\bibinfo {year} {2008})}\BibitemShut {NoStop}%
\bibitem [{\citenamefont {Calogero}(1969{\natexlab{a}})}]{calogero1969ground}%
  \BibitemOpen
  \bibfield  {author} {\bibinfo {author} {\bibfnamefont {F.}~\bibnamefont
  {Calogero}},\ }\href {\doibase 10.1063/1.1664821} {\bibfield  {journal}
  {\bibinfo  {journal} {J. Math. Phys.}\ }\textbf {\bibinfo {volume} {10}},\
  \bibinfo {pages} {2197} (\bibinfo {year} {1969}{\natexlab{a}})}\BibitemShut
  {NoStop}%
\bibitem [{\citenamefont
  {Calogero}(1969{\natexlab{b}})}]{calogero1969solution}%
  \BibitemOpen
  \bibfield  {author} {\bibinfo {author} {\bibfnamefont {F.}~\bibnamefont
  {Calogero}},\ }\href {\doibase 10.1063/1.1664820} {\bibfield  {journal}
  {\bibinfo  {journal} {J. Math. Phys.}\ }\textbf {\bibinfo {volume} {10}},\
  \bibinfo {pages} {2191} (\bibinfo {year} {1969}{\natexlab{b}})}\BibitemShut
  {NoStop}%
\bibitem [{\citenamefont {Calogero}(1971)}]{calogero1971solution}%
  \BibitemOpen
  \bibfield  {author} {\bibinfo {author} {\bibfnamefont {F.}~\bibnamefont
  {Calogero}},\ }\href {\doibase 10.1063/1.1665604} {\bibfield  {journal}
  {\bibinfo  {journal} {J. Math. Phys.}\ }\textbf {\bibinfo {volume} {12}},\
  \bibinfo {pages} {419} (\bibinfo {year} {1971})}\BibitemShut {NoStop}%
\bibitem [{\citenamefont {Calogero}(1975)}]{calogero1975exactly}%
  \BibitemOpen
  \bibfield  {author} {\bibinfo {author} {\bibfnamefont {F.}~\bibnamefont
  {Calogero}},\ }\href {\doibase 10.1007/BF02790495} {\bibfield  {journal}
  {\bibinfo  {journal} {Lett. Nuovo Cimento}\ }\textbf {\bibinfo {volume}
  {13}},\ \bibinfo {pages} {411} (\bibinfo {year} {1975})}\BibitemShut
  {NoStop}%
\bibitem [{\citenamefont {Sutherland}(1971)}]{sutherland1971quantum}%
  \BibitemOpen
  \bibfield  {author} {\bibinfo {author} {\bibfnamefont {B.}~\bibnamefont
  {Sutherland}},\ }\href {\doibase 10.1063/1.1665584} {\bibfield  {journal}
  {\bibinfo  {journal} {J. Math. Phys.}\ }\textbf {\bibinfo {volume} {12}},\
  \bibinfo {pages} {246} (\bibinfo {year} {1971})}\BibitemShut {NoStop}%
\bibitem [{\citenamefont {Polychronakos}(1992)}]{Polychronakos1992ExchangeOF}%
  \BibitemOpen
  \bibfield  {author} {\bibinfo {author} {\bibfnamefont {A.~P.}\ \bibnamefont
  {Polychronakos}},\ }\href {\doibase 10.1103/PhysRevLett.69.703} {\bibfield
  {journal} {\bibinfo  {journal} {Phys. Rev. Lett.}\ }\textbf {\bibinfo
  {volume} {69}},\ \bibinfo {pages} {703} (\bibinfo {year} {1992})}\BibitemShut
  {NoStop}%
\bibitem [{\citenamefont {Polychronakos}(2006)}]{Polychronakos_2006}%
  \BibitemOpen
  \bibfield  {author} {\bibinfo {author} {\bibfnamefont {A.~P.}\ \bibnamefont
  {Polychronakos}},\ }\href {\doibase 10.1088/0305-4470/39/41/s07} {\bibfield
  {journal} {\bibinfo  {journal} {J. Phys. A: Math. Gen.}\ }\textbf {\bibinfo
  {volume} {39}},\ \bibinfo {pages} {12793} (\bibinfo {year}
  {2006})}\BibitemShut {NoStop}%
\bibitem [{\citenamefont {Cirac}\ \emph {et~al.}(2011)\citenamefont {Cirac},
  \citenamefont {Poilblanc}, \citenamefont {Schuch},\ and\ \citenamefont
  {Verstraete}}]{PhysRevB.83.245134}%
  \BibitemOpen
  \bibfield  {author} {\bibinfo {author} {\bibfnamefont {J.~I.}\ \bibnamefont
  {Cirac}}, \bibinfo {author} {\bibfnamefont {D.}~\bibnamefont {Poilblanc}},
  \bibinfo {author} {\bibfnamefont {N.}~\bibnamefont {Schuch}}, \ and\ \bibinfo
  {author} {\bibfnamefont {F.}~\bibnamefont {Verstraete}},\ }\href {\doibase
  10.1103/PhysRevB.83.245134} {\bibfield  {journal} {\bibinfo  {journal} {Phys.
  Rev. B}\ }\textbf {\bibinfo {volume} {83}},\ \bibinfo {pages} {245134}
  (\bibinfo {year} {2011})}\BibitemShut {NoStop}%
\bibitem [{\citenamefont {Molignini}\ \emph {et~al.}(2022)\citenamefont
  {Molignini}, \citenamefont {L\'ev\^eque}, \citenamefont {Ke\ss{}ler},
  \citenamefont {Jaksch}, \citenamefont {Chitra},\ and\ \citenamefont
  {Lode}}]{CavityFermionization}%
  \BibitemOpen
  \bibfield  {author} {\bibinfo {author} {\bibfnamefont {P.}~\bibnamefont
  {Molignini}}, \bibinfo {author} {\bibfnamefont {C.}~\bibnamefont
  {L\'ev\^eque}}, \bibinfo {author} {\bibfnamefont {H.}~\bibnamefont
  {Ke\ss{}ler}}, \bibinfo {author} {\bibfnamefont {D.}~\bibnamefont {Jaksch}},
  \bibinfo {author} {\bibfnamefont {R.}~\bibnamefont {Chitra}}, \ and\ \bibinfo
  {author} {\bibfnamefont {A.~U.~J.}\ \bibnamefont {Lode}},\ }\href {\doibase
  10.1103/PhysRevA.106.L011701} {\bibfield  {journal} {\bibinfo  {journal}
  {Phys. Rev. A}\ }\textbf {\bibinfo {volume} {106}},\ \bibinfo {pages}
  {L011701} (\bibinfo {year} {2022})}\BibitemShut {NoStop}%
\bibitem [{\citenamefont {Lin}\ \emph {et~al.}(2019)\citenamefont {Lin},
  \citenamefont {Papariello}, \citenamefont {Molignini}, \citenamefont
  {Chitra},\ and\ \citenamefont {Lode}}]{PhysRevA.100.013611}%
  \BibitemOpen
  \bibfield  {author} {\bibinfo {author} {\bibfnamefont {R.}~\bibnamefont
  {Lin}}, \bibinfo {author} {\bibfnamefont {L.}~\bibnamefont {Papariello}},
  \bibinfo {author} {\bibfnamefont {P.}~\bibnamefont {Molignini}}, \bibinfo
  {author} {\bibfnamefont {R.}~\bibnamefont {Chitra}}, \ and\ \bibinfo {author}
  {\bibfnamefont {A.~U.~J.}\ \bibnamefont {Lode}},\ }\href {\doibase
  10.1103/PhysRevA.100.013611} {\bibfield  {journal} {\bibinfo  {journal}
  {Phys. Rev. A}\ }\textbf {\bibinfo {volume} {100}},\ \bibinfo {pages}
  {013611} (\bibinfo {year} {2019})}\BibitemShut {NoStop}%
\bibitem [{\citenamefont {Chomaz}\ \emph {et~al.}(2022)\citenamefont {Chomaz},
  \citenamefont {Ferrier-Barbut}, \citenamefont {Ferlaino}, \citenamefont
  {Laburthe-Tolra}, \citenamefont {Lev},\ and\ \citenamefont
  {Pfau}}]{Chomaz2022}%
  \BibitemOpen
  \bibfield  {author} {\bibinfo {author} {\bibfnamefont {L.}~\bibnamefont
  {Chomaz}}, \bibinfo {author} {\bibfnamefont {I.}~\bibnamefont
  {Ferrier-Barbut}}, \bibinfo {author} {\bibfnamefont {F.}~\bibnamefont
  {Ferlaino}}, \bibinfo {author} {\bibfnamefont {B.}~\bibnamefont
  {Laburthe-Tolra}}, \bibinfo {author} {\bibfnamefont {B.~L.}\ \bibnamefont
  {Lev}}, \ and\ \bibinfo {author} {\bibfnamefont {T.}~\bibnamefont {Pfau}},\
  }\href@noop {} {\enquote {\bibinfo {title} {Dipolar physics: A review of
  experiments with magnetic quantum gases},}\ } (\bibinfo {year} {2022}),\
  \Eprint {http://arxiv.org/abs/2201.02672} {arXiv:2201.02672} \BibitemShut
  {NoStop}%
\bibitem [{\citenamefont {Marcassa}\ and\ \citenamefont
  {Shaffer}(2014)}]{Marcassa2014}%
  \BibitemOpen
  \bibfield  {author} {\bibinfo {author} {\bibfnamefont {L.~G.}\ \bibnamefont
  {Marcassa}}\ and\ \bibinfo {author} {\bibfnamefont {J.~P.}\ \bibnamefont
  {Shaffer}},\ }\href {\doibase 10.1016/B978-0-12-800129-5.00002-X} {\bibfield
  {journal} {\bibinfo  {journal} {Adv. At. Mol. Opt. Phys.}\ }\textbf {\bibinfo
  {volume} {63}},\ \bibinfo {pages} {47} (\bibinfo {year} {2014})}\BibitemShut
  {NoStop}%
\bibitem [{\citenamefont {Chanda}\ \emph {et~al.}(2021)\citenamefont {Chanda},
  \citenamefont {Kraus}, \citenamefont {Morigi},\ and\ \citenamefont
  {Zakrzewski}}]{Chanda2021SelforganizedTI}%
  \BibitemOpen
  \bibfield  {author} {\bibinfo {author} {\bibfnamefont {T.}~\bibnamefont
  {Chanda}}, \bibinfo {author} {\bibfnamefont {R.}~\bibnamefont {Kraus}},
  \bibinfo {author} {\bibfnamefont {G.}~\bibnamefont {Morigi}}, \ and\ \bibinfo
  {author} {\bibfnamefont {J.}~\bibnamefont {Zakrzewski}},\ }\href {\doibase
  10.22331/q-2021-07-13-501} {\bibfield  {journal} {\bibinfo  {journal}
  {Quantum}\ }\textbf {\bibinfo {volume} {5}},\ \bibinfo {pages} {501}
  (\bibinfo {year} {2021})}\BibitemShut {NoStop}%
\bibitem [{\citenamefont {Chanda}\ \emph {et~al.}(2022)\citenamefont {Chanda},
  \citenamefont {Kraus}, \citenamefont {Zakrzewski},\ and\ \citenamefont
  {Morigi}}]{https://doi.org/10.48550/arxiv.2201.05466}%
  \BibitemOpen
  \bibfield  {author} {\bibinfo {author} {\bibfnamefont {T.}~\bibnamefont
  {Chanda}}, \bibinfo {author} {\bibfnamefont {R.}~\bibnamefont {Kraus}},
  \bibinfo {author} {\bibfnamefont {J.}~\bibnamefont {Zakrzewski}}, \ and\
  \bibinfo {author} {\bibfnamefont {G.}~\bibnamefont {Morigi}},\ }\href@noop {}
  {\enquote {\bibinfo {title} {Bond order via cavity-mediated interactions},}\
  } (\bibinfo {year} {2022}),\ \Eprint {http://arxiv.org/abs/2201.05466}
  {arXiv:2201.05466} \BibitemShut {NoStop}%
\end{thebibliography}%
\end{document}